\documentclass[aps,pra,twocolumn,floatfix,footinbib,notitlepage,showpacs]{revtex4-1}
\usepackage{graphicx,graphics,times,bm,bbm,bbold,amssymb,amsmath,amsfonts,dsfont,hyperref,color}
\usepackage[dvipsnames]{xcolor}
\usepackage{subfigure}
\hypersetup{colorlinks,linkcolor=blue,citecolor=blue,urlcolor=blue}

\newtheorem{theorem}{Theorem}

\usepackage[english]{babel}
\usepackage{natbib}
\usepackage[autostyle]{csquotes}
\usepackage{relsize}

\newcommand{\qed}{\nobreak \ifvmode \relax \else
	\ifdim\lastskip<1.5em \hskip-\lastskip
	\hskip1.5em plus0em minus0.5em \fi \nobreak
	\vrule height0.5em width0.60em depth0.2em\fi}

\newcommand{\ignore}[1]{}
\newcommand{\rd}{\mathrm{d}}

\newcommand{\re}{\mathrm{e}}

\newcommand{\rC}{\mathrm{C}}
\newcommand{\eref}[1]{(\ref{#1})}

\newcommand{\average}[1]{\left\langle#1\right\rangle}
\newcommand{\ket}[1]{|#1\rangle}
\newcommand{\cket}[1]{\left|#1\right)}
\newcommand{\braket}[2]{\left\langle#1|#2\right\rangle}
\newcommand{\cbraket}[2]{\left(#1|#2\right)}
\newcommand{\bra}[1]{\langle #1|}
\newcommand{\cbra}[1]{\left( #1\right|}

\newcommand{\qfi}[1]{\mathpzc{F}#1}
\newcommand{\ulc}[1]{\mathpzc{D}#1} 
\newcommand{\incom}[1]{\mathpzc{R}#1} 
\newcommand*{\rom}[1]{\expandafter\@slowromancap\romannumeral#1@}
\newcommand{\Tr}[1]{\mathrm{Tr}\left[#1\right]}

\newcommand{\mjd}[1]{\textcolor{red}{#1}}

\let\oldsqrt\sqrt
\def\sqrt{\mathpalette\DHLhksqrt}
\def\DHLhksqrt#1#2{%
\setbox0=\hbox{$#1\oldsqrt{#2\,}$}\dimen0=\ht0
\advance\dimen0-0.2\ht0
\setbox2=\hbox{\vrule height\ht0 depth -\dimen0}%
{\box0\lower0.4pt\box2}}
\DeclareFontFamily{OT1}{pzc}{}
\DeclareFontShape{OT1}{pzc}{m}{it}%
              {<-> s * [1.25] pzcmi7t}{}
\DeclareMathAlphabet{\mathpzc}{OT1}{pzc}%
                                 {m}{it}

\hypersetup{colorlinks,linkcolor={blue},citecolor={blue},urlcolor={blue}}
\begin{document}

\title{Multi-parameter quantum metrology with discrete-time quantum walks}
\author{Mostafa Annabestani}
\affiliation{Faculty of Physics, Shahrood University of Technology, Shahrood, Iran}
\email{Annabestani@shahroodut.ac.ir}
\author{Majid Hassani}
\affiliation{Department of Physics, Sharif University of Technology, Tehran 14588, Iran}
\email{corresponding author: majidhasani2010@gmail.com}
\author{Dario Tamascelli}
\affiliation{Quantum Technology Lab $\&$ Applied Quantum Mechanics Group, Dipartimento di Fisica ``Aldo Pontremoli'', Universit\`a degli Studi di Milano, I-20133 Milano, Italy}
\email{dario.tamascelli@unimi.it}
\author{Matteo G. A. Paris}
\affiliation{Quantum Technology Lab $\&$ Applied Quantum Mechanics Group, Dipartimento di Fisica ``Aldo Pontremoli'', Universit\`a degli Studi di Milano, I-20133 Milano, Italy}
\affiliation{INFN, Sezione di Milano, I-20133 Milano, Italy}
\email{matteo.paris@fisica.unimi.it}
\begin{abstract}
We address multi-parameter quantum estimation for one-dimensional discrete-time quantum walks and its applications to quantum metrology. We use the quantum walker as a probe for unknown parameters encoded on its coin degrees of freedom. We find an analytic expression of the quantum Fisher information matrix for the most general coin operator, and show that only two out of the three coin parameters can be accessed. We also prove that the resulting two-parameter coin model is asymptotically classical i.e. the Uhlmann curvature vanishes. Finally, we apply our findings to relevant case studies, including the simultaneous estimation of charge and mass in the discretized Dirac model. 
\end{abstract}
\date{\today}
\maketitle
\section{Introduction}
Quantum enhanced metrology ~\cite{Giovannetti2011advances,Paris2009quantum,Degen2017,Escher2011,
Braun2018} is among the most promising quantum technologies. Squeezing-enhanced optical interferometry~\cite{Caves1981,RafalPO} has been recently exploited in gravitational wave detectors~\cite{Acernese2019,Tse2019}, whereas quantum 
probes have carved their place into 
experimental investigation of delicate systems~\cite{Taylor2016}.
Several other applications of quantum enhanced sensors have been also suggested ~\cite{Budker2007,Koschorreck2010,Wasilewski2010,Sewell2012,Troiani2018,Ludlow2015,Louchet-Chauvet2010,Kessler2014a,Tamascelli2020,tama16}. Quantum metrology has its foundations in quantum estimation theory~\cite{Giovannetti2004Sci,Giovannetti2006prl,Giovannetti2011advances,Paris2009quantum,Degen2017,Escher2011, Demko2015,Polino2020,Hassani2017,majid19,seveso19}, which assesses the ultimate precision in the estimation of unknown parameters characterizing quantum systems and operations.

Since the early stages of quantum simulation, quantum walks have provided a formidable tool for both the determination of the computational power of quantum computers, and the study of {discrete quantum systems~\cite{feynman1985quantum,parthasarathy_1988,AharonovPRA,kempe2003,ChildsPRA}}. In fact, being the quantum analog of classical random walks (on configuration spaces), quantum walks in either the discrete- or continuous-time version~\cite{Farhi98PRA} provide a simple but powerful instrument to define and characterize quantum algorithms~\cite{ambainis2007} and communication protocols in the same way as classical random walks are used to analyze randomized algorithms. On the other side, discrete time and space version of fundamental equations, such as the Dirac equation, can be interpreted as coined {discrete-time quantum walks on lattices~\cite{Chandrashekar10PRA,Arrighi2014,Chandrashekar16SR,Arnault16PA,Mallick17,Arrighi18PRA,Arnault19PRA,DeVincenzo19RP}}. In view of their use for in the realization of quantum protocols, the characterization of quantum walks at the quantum level is a necessary step. Indeed, quantum metrological schemes have been proposed and precision benchmarks have been obtained for  either {the discrete~\cite{Singh}- or continuous-time quantum walks~\cite{Zatelli20}}.

In this work, we assess the ultimate precision attainable in the determination of the unknown coin operator in a discrete-time quantum walk. Since three parameters are necessary, in the general setting, to define a unitary operator acting on a two-level system, multi-parameter quantum estimation theory will be exploited to determine bounds on the efficiency of unbiased estimators of the coin operator. 
As we will see by exploiting our analytic expression of the quantum Fisher information matrix, only two out of the three parameters defining the coin operator may be actually accessed. On the other hand, we prove that for the resulting two-parameter coin model there is no incompatibility, i.e. the Uhlmann curvature vanishes and the model is asymptotically classical. {This fact implies that the quatum walker can be used as an optimal probe in the multi-parameter quantum metrology which yields to the compatible model of estimation~\cite{Ragy2016}.
}

The paper is organized as follows. We will first briefly introduce multi-parameter quantum estimation and define the relevant quantities in {Section \ref{Int.Multi}. Section \ref{DTQW}} is devoted to coined discrete-time quantum walks. In {Section \ref{full-part}} we present  our main results about the analytic expression of the quantum Fisher information matrix and Uhlmann curvature (or incompatibility) matrix. We proceed in {Section \ref{Examples}} with presenting some relevant examples, that allow us to establish the compatibility of our findings with what already present in the literature, and to illustrate the scope and range of our results. {Section \ref{Conclusion}} closes the paper with some concluding remarks.
\section{multi-parameter quantum estimation}\label{Int.Multi}
Quantum estimation theory deals with the assessment of the ultimate estimation precision attainable in the presence of quantum resources, such as coherence and entanglement. The estimation process can be ideally seen as follows: a quantum system, or probe, is prepared in a particular initial state, evolved under the action of an Hamiltonian, or Liouvillian, having unknown parameters, and then measured. The goal of this procedure is to gain as much information as possible on the unknown parameters, which {determine} the dynamics of  the probe. For fixed initial state and evolution, the amount of information that is accessed upon measurement depends on the measurement itself, whose outcome is then suitably processed by means of an estimator. Analogously to the classical Cram{\`e}r-Rao bound, limiting from above the efficiency of estimators in terms of a quantity independent of the estimator itself, i.e. the Fisher information... (Graybill), the quantum Cram{\`e}r-Rao bound fixes an upper bound to the  the ultimate efficiency of the parameter estimation in the presence of quantum resources. In the case of a single unknown parameter $\theta$, once indicated by $\{\rho_\theta\}$ the family of quantum states depending on $\theta$ the ultimate precision of any unbiased estimator {$\widehat{\theta}$} for $\theta$ is given by quantum Cram{\`e}r-Rao inequality:
\begin{equation} \label{eq:classicalCRB}
\sigma^2[\widehat{\theta}]\geq \frac{1}{\qfi(\theta)},
\end{equation}
where  $\sigma^2$ is the variance of the estimator and $F(\theta)$ is the quantum Fisher information (QFI) defined as
\begin{equation} \label{eq:QFI1}
\qfi(\theta)=\Tr{\rho_{\theta} L_{\theta}^2 }.
\end{equation}
 $L_{\theta}$ is the symmetric logarithmic derivative (SLD) implicitly defined by  
 \begin{equation} \label{eq:symmLog1}
 \frac{\partial\rho_{\theta}}{\partial\theta} = \frac12 \{L_{\theta},\rho_{\theta}\},
 \end{equation}
and $\{\cdot\}$ denotes the anti-commutator. The QFI is therefore, as its classical counterpart, independent of the measurement. We moreover remark that {the optimal measurement}, i.e. the one saturating the quantum Cram\'er-Rao inequality, is the projector operator which can be constructed by the eigenvectors of the SLD~\cite{Paris2009quantum}.

  These results and definitions can be extended to the multi-parameter estimation scenario{~\cite{Holevo1973,Holevo1977,Helstrom1976,Holevo2011,Helstrom1967pla}}, namely when the unknown parameters to be jointly estimated are more than one. In this case we indicate by $\{\rho_\Theta\}$ the statistical model, with $\Theta = (\theta_1,\theta_2,\ldots,\theta_n)$ and $\theta_i \in  \mathbb{R}$. The Cram\'er-Rao bound \eref{eq:classicalCRB} for multi-parameter estimation is expressed as \cite{Hayashi2008}

\begin{equation}\label{CRB}
\text{Cov}(\Theta)\geqslant\qfi ^{-1},
\end{equation}
where $ \text{Cov}(\Theta) $ is the $n \times n$ covariance matrix; $ \qfi $ denotes instead the quantum information Fisher matrix (QFIm) with elements defined as
\begin{equation}\label{mqfi}
\qfi _{\mu\nu}=\frac{1}{2}\Tr{\varrho _{\Theta}\lbrace L_{\mu},L_{\nu}\rbrace},
\end{equation}
where  $ L_{\mu} $ denotes the SLD with respect to the $\mu$-th parameter. From now on we omit the $ \Theta $ subscript wherever clear from the context for ease of notation. 

 Differently from the single-parameter case, in the multi-parameter scenario the quantum limit given by the matrix inequality  \eref{CRB} is not achievable in general~\cite{Baumgratz2016,Carollo2019,Albarelli2019prl,Albarelli2019,Demko2020,Ragy2016,Razavian2020}. This fact has its root in the non-commuting nature of the operator algebra, preventing the simultaneous measurement of arbitrary observables with arbitrary accuracy and leading to trade offs for the precision of the individual estimators.
 
 A most useful scalar bound can be obtained by introducing a real and positive weight matrix $ W $. This yields
\begin{equation}\label{CRB-sym}
\Tr{\text{Cov}(\Theta)W}\geqslant\rC ^{S}(\Theta , W),
\end{equation}
where
\begin{equation}\label{sym-bound}
\rC ^{S}(\Theta , W)=\Tr{\qfi ^{-1}W},
\end{equation}
also known as the symmetric bound. A tighter scalar bound was derived by Holevo \cite{Holevo1973,Holevo2011}, which can be numerically calculated by means of linear semi-definite programming as \cite{Albarelli2019prl}. Recently, it has been proved that the Holevo bound $\rC ^{H}(\Theta , W)$ can be upper bounded by the symmetric bound as follows \cite{Carollo2019}
\begin{equation} \label{eq:symmHolevo}
\rC ^{S}(\Theta , W)\leqslant\rC ^{H}(\Theta , W)\leqslant(1+\incom)\rC ^{S}(\Theta , W).
\end{equation}
The quantity $\incom$ is defined as
\begin{equation}\label{R-def}
\incom =\big\Vert i\qfi ^{-1}\ulc\big\Vert _{\infty},
\end{equation}
where $ \Vert A\Vert _{\infty}$ is the largest eigenvalue of the matrix $A$, and $0 \leq  \incom \leq 1$. The coefficient $\incom$ measures the amount of incompatibility of the unknown parameters, and is in fact defined in terms of the Uhlmann curvature matrix  
\begin{equation}\label{ulc}
\ulc _{\mu\nu}=-\frac{i}{2}\Tr{\varrho _{\Theta}\left[  L_{\mu},L_{\nu}\right] }.
\end{equation}

{Strictly speaking, multi-parameter quantum metrology corresponds to simultaneous 
estimation of multiple parameters using a single quantum system to probe a quantum 
dynamics with unknown parameters. Of course, separate experiments may be also exploited, and in this case each parameter is independently estimated. This means that in every estimation run all parameters except one are considered perfectly known. The symmetric bound is not generally achievable in the simultaneous estimation, corresponding to the fact that 
in simultaneous estimation one uses only the resources of one of the separate schemes. 

{The compatibility conditions are as follows: i) The Uhlmann curvature matrix vanishes, $ \ulc _{\mu\nu}=0 $; this requirement ensures the existence of compatible measurements and the saturability of the symmetric bound, ii) The QFIm is a diagonal matrix, i.e. $ \qfi _{\mu\nu}=0 $ for all $ \mu\neq\nu $, which implies that the different parameters can be estimated independently, and iii) There exists a single probe state that maximizes the QFIs for all parameters. When these compatibility conditions are fulfilled, the performance of the simultaneous and separate schemes will be equal to each other. Such models are referred to as \textit{compatible} models~\cite{Ragy2016}. In compatible models each of parameters is estimated independently with ultimate precision, and typically require fewer resources with respect to separate schemes~\cite{Ragy2016,FidererPRX}.} In this work, we prove that discrete-time quantum walks provide a compatible model for multi-parameter quantum metrology.}

\section{Discrete-time quantum walk}\label{DTQW}
In a discrete-time quantum walk (DTQW) on a one dimensional lattice, at each time step a walker moves between nearest-neighbor sites of the lattice with an amplitude that depends on the state of a two-level system playing the role of a coin\mjd{~\cite{AharonovPRA,kempe2003}}. Between different moves, moreover, the state of the coin can be modified by some operator. The Hilbert space  $ \mathcal{H}_{p} $ of the walker is spanned by the elements of the position basis $ \lbrace \ket{x}\mid x\in\mathbb{Z}\rbrace$, where $\ket{x}$ indicates that the walker is on the $x$-th site of the lattice. A basis for the space $ \mathcal{H}_{c} $ of the coin is instead provided by the eigenstates  $ \lbrace\ket{0},\ket{1}\rbrace $ of the Pauli matrix $\sigma_z$, and the complete lattice-coin space is  $\mathcal{H} = \mathcal{H}_p \otimes \mathcal{H}_c$.

In this basis, the evolution of the quantum walk is therefore determined by the repeated application to an initial state of the form 
\begin{equation}
     \ket{\Psi (0)}=\sum\nolimits_{x,j}{c_{x,j}(0)\ket{x}_{p} \otimes\ket {j}_{c}}
\end{equation}
of the operator
\begin{equation}
    U_\Theta = S (\mathbb{1}\otimes C_\Theta),
\end{equation}
where
\begin{equation}\label{shift-X}
S=\sum_{x}{\ket{x+1}\bra{x}\otimes\ket{0}\bra{0}+\ket{x-1}\bra{x}\otimes\ket{1}\bra{1}}
\end{equation}
is the conditional shift operator and $C_\Theta$ is the coin operator.
Neglecting an overall phase factor the most general form of the coin operator, i.e. an element of $U(2)$, is given by
\begin{equation} \label{CoinOperator}
C_{\Theta}=C_{\theta,\alpha,\beta}=\left(\begin{array}{cc}
\re ^{i\alpha}\cos\theta &\re ^{i\beta}\sin\theta\\ 
-\re ^{-i\beta}\sin\theta &\re ^{-i\alpha}\cos\theta\
\end{array}\right).
\end{equation}
The parameters $\theta,\alpha$ and $\beta$ are the unknown parameters addressed by our multi-parameter estimation problem.
The  state of the quantum walker after $ t $ steps is equal to 
\begin{equation}\label{finalstate}
\ket{\Psi _{\Theta}(t)}=U_{\Theta}^{t}\ket{\Psi (0)}.
\end{equation}
For our purposes it is expedient to work with a diagonal representation of the shift operator. To this end we define a new basis for the position space by applying the  Fourier transformation~{\cite{nayak2000}}

\begin{equation}\label{Fourier}
\ket{k}=\sum _{x}\re ^{ikx}\ket{x}.
\end{equation}
The set  $\{ \ket{k} \},\ -\pi \leq k \leq \pi $ satisfies the completeness condition
\begin{align}
\frac{1}{2\pi}\int _{-\pi}^{\pi}\rd k~\ket{k}\bra{k}= \mathbb{1},\label{completeness}
\end{align} 
and, moreover,
\begin{align}
\delta(k-k')=\frac{1}{2\pi}\sum _{x}\re ^{-i(k-k')x}.\label{deltafun}
\end{align}

The unitary operator in $ k $-space is given by 
\begin{equation}\label{U-Uk}
U_{\Theta}=\frac{1}{2\pi}\int _{-\pi}^{\pi}\rd k~\ket{k}\bra{k}\otimes {u}_{k}(\Theta),
\end{equation}
with 
\begin{align}
{u}_{k}(\Theta)&=\left(\re ^{-ik}\ket{0}\bra{0}+\re ^{ik}\ket{1}\bra{1}\right)C_{\Theta}.\label{Uk}
\end{align}
Replacing Eq. (\ref{U-Uk}) in Eq. (\ref{finalstate}) yields
\begin{equation}\label{finalkspace}
\ket{\Psi _{\Theta}(t)}=\frac{1}{2\pi}\int _{-\pi}^{\pi}\rd k~\ket{k}\otimes\ket{\varphi _{k}^{t}(\Theta)},
\end{equation}
where $ \ket{\varphi _{k}^{t}(\Theta)}={u}_{k}^{t}\ket{\varphi_k\left(0\right)}$ and
\begin{align}\label{initial-k}
\ket{\varphi_k\left(0\right)}=\braket{k}{\Psi (0)},
\end{align}	
is the amplitude of initial state in $k$-space. We observe that we have replaced ${u}_{k}^{t}(\Theta)$ by ${u}_{k}^{t}$ to simplify the notation. It is clear from Eq. (\ref{finalkspace}) that in the $k$-space the state of the walker after $ t $ steps will be block-diagonal and that the parameter dependence appears only in the coin part.

\section{Main result}\label{full-part}
Equation (\ref{finalkspace}) is the pure state of DTQW after $ t $ steps which is defined on the whole Hilbert space, namely the coin and the  position space. By exploiting the fact that for pure states the relations $ \left( \varrho^{2}=\varrho =\ket{\Psi }\bra{\Psi }\right) $ and $ L_{\mu}=2\partial _{\mu}\varrho=2\left( \ket{\partial _{\mu}\Psi }\bra{\Psi }+\ket{\Psi }\bra{\partial_{\mu}\Psi }\right)$ hold, it is possible to determine the QFIm and the Uhlmann curvature matrix; their elements are given by 
\begin{align}
\qfi _{\mu\nu}[\ket{\Psi }\bra{\Psi }]&=4\,\mathfrak{R}\left(\average{\partial _{\mu}\Psi\Big\vert\partial _{\nu}\Psi}-\average{\partial _{\mu}\Psi\Big\vert\Psi}\average{\Psi\Big\vert\partial _{\nu}\Psi} \right)\label{qfi-pure},\\
\ulc _{\mu\nu}[\ket{\Psi }\bra{\Psi }]&=4\,\mathfrak{I}\left(\average{\partial _{\mu}\Psi\Big\vert\partial _{\nu}\Psi}-\average{\partial _{\mu}\Psi\Big\vert\Psi}\average{\Psi\Big\vert\partial _{\nu}\Psi} \right)\label{ulc-pure},
\end{align}
where $ \mathfrak{R} $ and $ \mathfrak{I} $ denote, respectively, the real and imaginary part, and  $\partial _{\mu}= \frac{\partial}{\partial\theta _{\mu}}$. The first derivative of Eq. (\ref{finalkspace}) is 
\begin{align}
\ket{\partial _{\mu}\Psi _{\Theta}(t)}&=\frac{1}{2\pi}\int _{-\pi}^{\pi}\rd k~\ket{k}\otimes\ket{\partial _{\mu}\varphi _{k}^{t}(\Theta)}
\end{align}
where
\begin{align}
\ket{\partial _{\mu}\varphi _{k}^{t}(\Theta)}&=\sum _{m=0}^{t-1} {u}_{k}^{m+1}\,O_{\mu}\, {u}_{k}^{m+1\,\dagger}\ket{\varphi _{k}^{t}(\Theta)},\label{1deriv-coinpart}
\end{align}
and $ O_{\mu}= {u}_{k}^{\,\dagger}\,\partial _{\mu} {u}_{k} $ (see Appendix \ref{appendix-1deriv-coinpart} for details on the derivation). One can define the \textit{superoperator}, $ \mathcal{A}_{k} $ on the coin space of the walker as 
\begin{align}
\sum _{m=0}^{t-1} {u}_{k}^{m +1}\,O_{\mu}\, {u}_{k}^{m+1\,\dagger}&\equiv\sum_{m =0}^{t-1}\mathcal{A}_{k}^{m +1}({O}_{\mu})=\mathcal{A}_{k}^{'}({O}_{\mu}).\label{superop-deriv}
\end{align}
This yields 
\begin{align}\label{pure-qrw-fund}
&\average{\partial _{\mu}\Psi\Big\vert\partial _{\nu}\Psi}-\average{\partial _{\mu}\Psi\Big\vert\Psi}\average{\Psi\Big\vert\partial _{\nu}\Psi}\\
&=\int _{-\pi}^{\pi}\frac{\rd k}{2\pi}\, \average{\mathcal{A}_{k}^{'}({O}_{\mu}^{\dagger})\mathcal{A}_{k}^{'}({O}_{\nu})}_{t}- \nonumber\\
&\left(\int _{-\pi}^{\pi}\frac{\rd k}{2\pi}\, \average{\mathcal{A}_{k}^{'}({O}_{\mu}^{\dagger})}_{t} \right)\left(\int _{-\pi}^{\pi}\frac{\rd k}{2\pi}\, \average{\mathcal{A}_{k}^{'}({O}_{\nu})}_{t} \right)\nonumber,
\end{align}
where
\begin{align}
\average{\bullet} _{t}&=\bra{\varphi _{k}^{t}(\Theta)} \bullet\ket{\varphi _{k}^{t}(\Theta)}=\Tr{\bullet\,\mathcal{A}_{k}^{t}(\varrho_0)}\label{def-expect},
\end{align}
with $\varrho_0=\ket{\varphi_k\left(0\right)}\bra{\varphi_k\left(0\right)}$. In order to extract simple analytic relations for $ \qfi _{\mu\nu} $ and $ \ulc _{\mu\nu} $, we adopt the {superoperator formalism}\cite{Brun2003,Mostafa2010}.    

 Any two-dimensional Hermitian (anti-Hermitian) operator like $O$ can be represented in terms of Pauli matrices $\{ \mathbb{1},\sigma _{x},\sigma _{y},\sigma _{z}\}$ as follows
\begin{equation}\label{O-pauli}
O=\frac{1}{2}\left( o_0 \mathbb{1}+o_{x} \sigma _{x}+o_{y} \sigma _{y}+o_{z} \sigma _{z}\right) ,
\end{equation}
where the coefficients $o_i$ are determined by the Hilbert-Schmidt product $\Tr{O e_i}$ of $O$ with the $i$-th element of a basis for the space of $2\times 2$ matrices. We set $o_i=\Tr{O\sigma _i},\, i\in \{0,x,y,z\}$, and $ \sigma _{0}= \mathbb{1} $. The coefficients of the above expansion can be regarded as the elements of four-dimensional column vector 
\begin{equation}\label{4DVRofO}
\cket{O}=\left(\begin{array}{cccc}
o_{0}\\ 
o_{x}\\
o_{y}\\
o_{z}\\
\end{array}\right)\equiv \left(o_{0},\vec{o}\right)^{T},
\end{equation}
in which $ \vec{o} $ is nothing else than the \textit{Bloch vector}. The Bloch vector reminiscent of the Bloch representation for any density matrix.
 
\begin{theorem}\label{theorem1}
The elements of the QFIm $\qfi_{\mu\nu}$  and of the Uhlmann curvature matrix $\ulc_{\mu\nu}$ 
are
\begin{align}\label{pure-qfim-theorem}
\qfi _{\mu\nu}&=t^{2}\left\lbrace\int _{-\pi}^{\pi}\frac{\rd k}{2\pi}\, \cbra{O_\mu}\mathcal{A}_{k}^{\mathbb{1}}\cket{O_\nu} \right . \\
&\left . - \left(\int _{-\pi}^{\pi}\frac{\rd k}{2\pi}\, \cbra{O_\mu}\mathcal{A}_{k}^{\mathbb{1}}\cket{ \varrho_0} \right)\left(\int _{-\pi}^{\pi}\frac{\rd k}{2\pi}\, \cbra{ \varrho_0}\mathcal{A}_{k}^{\mathbb{1}}\cket{O_\nu} \right)\right\rbrace , \nonumber \\
\ulc_{\mu\nu}&=0,
\end{align}
where $O_i= {u}_{k}^{\,\dagger}\,\partial _{i} {u}_{k} ,\, i\in\lbrace \alpha,\beta,\theta \rbrace$, $u_k$ represents the evolution operator, $\varrho_0=\ket{\varphi_k\left(0\right)}\bra{\varphi_k\left(0\right)}$ denotes the initial state of the DTQW in $k$-space, and
\begin{widetext}
\begin{align}\label{A1-Explicit}
\mathcal{A}_{k}^{\mathbb{1}}=N\left( \begin {array}{cccc} \frac{1}{N}&0&0&0\\ \noalign{\medskip}0&{
 \sin ^{2}\left( k-\beta \right) 
 }&{- \cos \left( k-\beta \right) \sin \left( k-\beta
\right) }&{\cot \theta \sin
\left( k-\beta \right) \sin \left( k-\alpha \right) }\\ \noalign{\medskip}0&{-\cos \left( k-\beta \right) \sin \left( k-\beta
\right) }&{\cos ^{2}\left( k-\beta \right)}&{-\cot\theta\cos \left( k-
\beta \right) \sin \left( k-\alpha \right)}\\ \noalign{\medskip}0&{\cot \theta \sin
\left( k-\beta \right) \sin \left( k-\alpha \right) }&{-\cot\theta\cos \left( k-
\beta \right) \sin \left( k-\alpha \right)}&{\cot^2 \theta\sin ^{2}\left( k-\alpha
\right)}
\end {array} \right),
\end{align}
\end{widetext}

with $N=\frac{\sin^2\theta}{1-\cos^2\theta\cos^2\left(k-\alpha\right)}$.
\end{theorem}
For the sake of readability, the proof of the theorem, requiring quite heavy notation, is presented in Appendix  \ref{theorem1Proof}. Here we limit ourselves to remark some consequences of our main result.

\textbf{Corollary 1:} Only the parameters $\theta$ and $\alpha$ of the coin operator can be estimated.

\textbf{Proof:} By exploiting the definition $O_i= {u}_{k}^{\,\dagger}\,\partial _{i} {u}_{k}$, one can easily show that
\begin{equation}\label{eq:OEthetaxplicit}
\cket{O_{\theta}}=2i\left(\begin{array}{cccc}
0\\ 
-\sin (\alpha -\beta)\\
\cos (\alpha -\beta)\\
0\\
\end{array}\right),
\end{equation}

\begin{equation}\label{eq:OEalphaxplicit}
\cket{O_{\alpha}}=i\left(\begin{array}{cccc}
0\\ 
\cos (\alpha -\beta)\,\sin 2\theta\\
\sin (\alpha -\beta)\,\sin 2\theta\\
2\cos ^{2}\theta\\
\end{array}\right),
\end{equation}
\begin{equation} \label{eq:OEbetaexplicit}
\cket{O_{\beta}}=i\left(\begin{array}{cccc}
0\\ 
\cos (\alpha -\beta)\,\sin 2\theta\\
\sin (\alpha -\beta)\,\sin 2\theta\\
-2\sin ^{2}\theta\\
\end{array}\right).
\end{equation}
It is thus straightforward to show that
\begin{align}\label{AbetaZero}
\mathcal{A}_{k}^{\mathbb{1}}\cket{O_\beta}=0.
\end{align}
{This result implies that $ \qfi_{\mu\beta}=0 $ for $ \mu\in\lbrace\theta,\alpha\rbrace $; all nonvanishing elements of $ \qfi_{\mu\nu} $ are therefore given by $ \theta $ and $ \alpha $.}

\textbf{Corollary 2:}\label{corollary2} {By vanishing the second term of the $ \qfi_{\mu\nu} $ in Eq.~(\ref{pure-qfim-theorem}) which, in contrast to first one}, is initial-state dependent, the maximum value of the diagonal elements of the QFIm is given by
\begin{align}\label{pure-qfim-theorem-maximum}
\qfi _{\mu\mu}&=t^{2}\int _{-\pi}^{\pi}\frac{\rd k}{2\pi}\, \cbra{O_\mu}\mathcal{A}_{k}^{\mathbb{1}}\cket{O_\mu} \\
&=\left\{ {\begin{array}{*{20}{c}}
{{4t^2\sin\theta}\left(1+\sin\theta\right)^{-1}\quad\mu=\theta},\\
{4t^2\left(1-\sin\theta\right)\qquad\quad\quad{\mu=\alpha}},
\end{array}} \right. \nonumber
\end{align}
which explicitly depends the  single parameter $ \theta $. Moreover, if this limit is saturated by a suitable choice of the initial state, then the QFIm will be diagonal. This is due the form of the off-diagonal elements ($\mu\neq\nu$), which contain the initial-state independent terms
\begin{align}
\int _{-\pi}^{\pi}\frac{\rd k}{2\pi}\, \cbra{O_{\mu}}\mathcal{A}_{k}^{\mathbb{1}}\cket{O_{\nu}}=0.
\end{align}

\textbf{Corollary 3:} In order to simultaneously maximize both  the diagonal elements $\qfi _{\theta\theta}$ and $\qfi _{\alpha\alpha}$ of the QFIm--see Corollary 2, one should solve $\qfi _{\theta\theta}=\qfi _{\alpha\alpha}$ which admits solutions  $\sin\theta=-g$ and $\sin \theta = -g^{-1}$, where $g=\frac{1+\sqrt{5}}{2}$ and $g^{-1}=\frac{1-\sqrt{5}}{2}$ are the ``golden ratio'' and the ``golden ratio conjugate'', respectively. Since $g>0$, we have $\theta=\arcsin(-g^{-1})$.

\textbf{Corollary 4:}
If the walker is initially localized on a site of the lattice, the elements of the QFIm are explicitly given by 
\begin{align}
\qfi _{\theta\theta}&=\frac{4t^{2}}{1+\sin\theta}\left(\sin\theta-\frac{\left( \hat{n}\times\vec{r}\right)_{\hat{z}}^{2}}{1+\sin\theta}\right),\label{qfi-tt-full}\\
\qfi _{\phi\phi}&= 4t^{2}\left(1-\sin\theta\right) \left(1-\frac{\left(\hat{n} \cdot \vec{r}\right)^{2}}{1+\sin\theta}\right),\label{qfi-aa-full}\\
\qfi_{\theta\phi}&=\frac{-4t^{2}\left(1-\sin\theta\right)}{\cos\theta\left( 1+\sin\theta\right)}\left(\hat{n} \cdot \vec{r}\right)\left( \hat{n}\times\vec{r}\right)_{\hat{z}},\label{qfi-ta-full}
\end{align}
where $\phi=\alpha-\beta$, $\hat{n}=\left(\sin\theta\cos\phi, \sin\theta\sin\phi, \cos\theta \right)$, $\vec{r}$ is the Bloch vector of the coin initial state, and $( \cdot)_{\hat{z}}$ indicates the third component of the vector. {In particular, for a localized initial state, the diagonal form of the QFIm is attainable by choosing $\vec{r}=\partial_{\theta}\hat{n}$---see Corollary 2.}

\textbf{Proof:} All of $\cket{O_i}$s depend on $\theta$ and $\phi=\alpha-\beta$ see Eqns. \eref{eq:OEthetaxplicit}-\eref{eq:OEbetaexplicit}.  In addition, for any local initial state in the position space
\begin{align}\label{local-initial-state}
\ket{\Psi(0)}=\ket{x_{0}}_{p}\otimes\ket{\chi}_{c},
\end{align}
and $\varrho_0=\ket{\varphi_k\left(0\right)}\bra{\varphi_k\left(0\right)}=\ket{\chi}\bra{\chi}$ does not depend on $k$. The integration in Eq. (\ref{pure-qfim-theorem}) is thus taken over $\mathcal{A}_{k}^{\mathbb{1}}$ and the solution of the integral depends on $\theta$ and $\phi$.  It follows that the elements of the QFIm have only two independent parameters ($\theta$ and $\phi$). Equations (\ref{qfi-tt-full}), (\ref{qfi-aa-full}), and (\ref{qfi-ta-full}) can be easily derived by calculating Eq. (\ref{pure-qfim-theorem}) with $ \varrho_0\equiv\cket{ \varrho_0}=\left(1,\vec{r}\right)^T$.

{\begin{theorem}\label{theorem2}
The choice of an optimal initial state of the quantum walker, namely an initial state that makes the initial state dependence expressed by the second term of the QFIm \eref{pure-qfim-theorem}, the DTQW yields to a compatible model in the multi-parameter quantum metrology.
\end{theorem}}

{\textbf{Proof:} Theorem~\ref{theorem1} and Corollary 2 indicate that choosing the optimal initial states which maximize the both diagonal elements of the QFIm and vanish the off-diagonal elements of that, satisfy three compatibility conditions.} 

{Let consider the initial-state dependent term of $ \qfi $ in Eq.~(\ref{pure-qfim-theorem}) 
\begin{equation}\label{Intialstatedependent}
\int _{-\pi}^{\pi}\frac{\rd k}{2\pi}\, \cbra{O_{\mu}}\mathcal{A}_{k}^{\mathbb{1}}\cket{\varrho ' _{0}},~~~~\mu\in\lbrace\theta, \alpha\rbrace .
\end{equation}
One can suppress this term by two approaches: i) choosing initial states which satisfy $ \mathcal{A}_{k}^{\mathbb{1}}\cket{\varrho ' _{0}}=0 $; ii) choosing an entangled initial state between the position 
space and the coin space. In particular, we observe that for any entangled initial state of the walker of 
the  form
\begin{equation}\label{entangledInitial}
\ket{\Psi ' (0)}=\frac{1}{\sqrt{2}}\left(\ket{x_{1,p}0_{c}}+\ket{x_{2,p}1_{c}} \right),
\end{equation}
where $ x_{1} $ and $ x_{2} $ indicate two points in the position space and $\vert x_{1}-x_{2}\vert $ is an \textit{odd number}, we have
\begin{equation*}
\int _{-\pi}^{\pi}\frac{\rd k}{2\pi}\, \cbra{O_{\mu}}\mathcal{A}_{k}^{\mathbb{1}}\cket{\varrho ' _{0}}=0,~~~~\mu\in\lbrace\theta, \alpha\rbrace .
\end{equation*}
In this case $ \cket{\varrho ' _{0}} $ do depend on $ k $ but does not depend on the unknown parameters. Hence by these two approaches, one can find the optimal initial state of the quantum walker which assures, together with the fulfillment of the other conditions, the compatibility.
}

{A  (discrete-time) quantum walker can be used as a probe in a multi-parameter quantum metrology scenario in which the unknown parameters are encoded on the coin space of the walker. The evolution is governed by the unitary operation of the DTQW. As we showed, by  suitably choosing the  initial state, the DTQW yields to a compatible model, where all parameters are estimated with ultimate precision independently while consuming fewer resources. Moreover, optimal entangled initial states does not depend to the unknown parameters, in contrast to optimal local initial states---see Corollary 4. We illustrate some relevant applications of our results by means of the following examples.
}
\section{Case studies}\label{Examples}
In order to gain insight about the applications of the above general results, let us now consider few examples. In particular, we aim to show the applicability of our results in DTWQ-related quantum metrology problems.

\subsubsection{Single-parameter quantum metrology}
\begin{figure} 
\includegraphics[width=8.5cm]{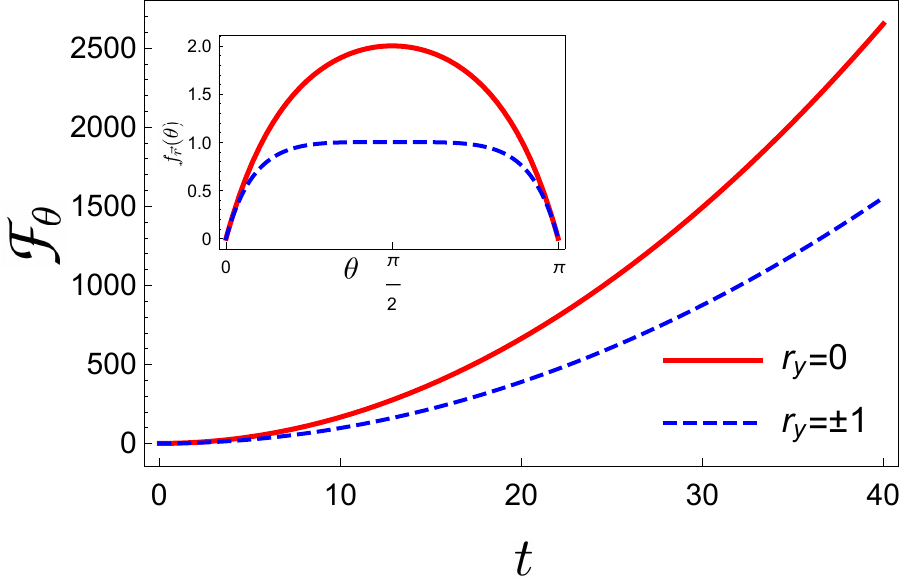}
\caption{{
The QFI \eref{eq:QFI1Tama} for $\theta=\pi/4$ as a function of time when a maximizing ($r_y=0$, solid red line) or a minimizing (eigenstates of $\sigma_y$, dashed blue line) initial state is selected. In the inset the prefactor $f_{\vec{r}}(\theta)$ for the same maximizing (solid red line)  and minimizing (dashed blue line) initial states as a function of $\theta$.}} 
\label{Single-Fisher-Tot} 
\end{figure}

{We start by appling our results to the single-parameter case. Without loss of generality in what follows we set  $\alpha = \beta = 0$, and focus on the remaining parameter $\theta$. In this case Eq.\eref{qfi-tt-full} simplifies to }
\begin{equation} \label{eq:QFI1Tama}
 \qfi _{\theta} = t^2 f_{\vec{r}}(\theta) = t^2 \frac{4  \sin \theta \left [1+\sin \theta\left (1-r_y^2) \right )\right ]}{(1+\sin \theta)^2} .   
\end{equation}
{The ultimate estimation accuracy, as determined by the QFI, is thus monotonically increasing with time; for fixed value of the unknown parameter $\theta$, moreover, the QFI is a function of the initial state and is maximized by the set of states lying in the $x-z$ plane for which $r_y = 0$, whereas the $\pm1$ eigenstates of $\sigma_y$ minimize the QFI. As shown in Fig.\ref{Single-Fisher-Tot} at any fixed time the difference between the optimal and worse choice of the initial state can lead to $1/\sqrt 2$ factor in the standard deviation of the efficient estimator of $\theta$.}

The minimum value of the $ \qfi_{ \theta} $ is compatible with the results in Ref.\cite{Singh}, where numerical evaluation for each initial state of the coin was required.

\subsubsection{Two-parameter quantum metrology with special initial states of the coin}
\begin{figure}
\includegraphics[width=8.5cm]{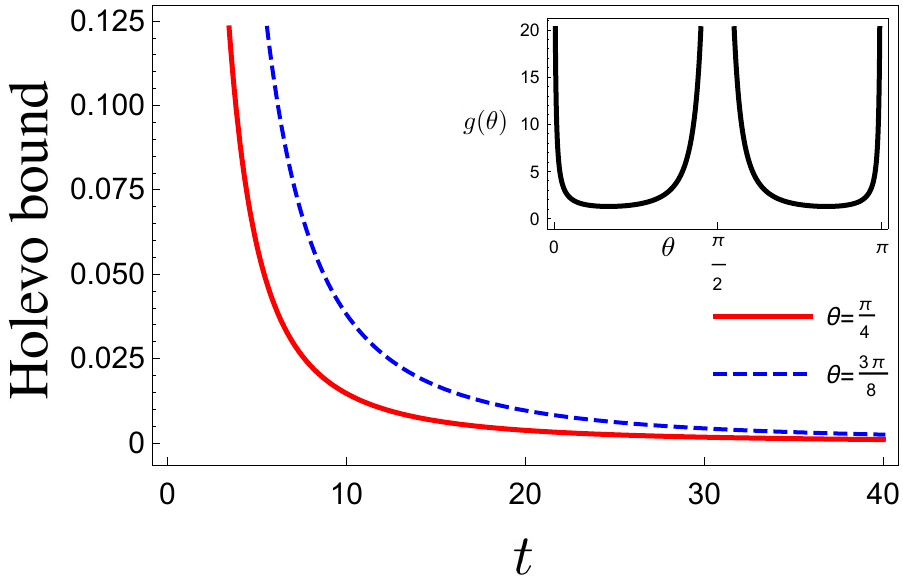}
\caption{{The behavior of the Holevo bound for the full state of the walker as a function of time for $ \theta=\pi/4 $ and $ \theta=3\pi/8 $. In the inset, the quantity $f_{\vec{r}}$ (see Eq.\eref{eq:QFI1Tama}) as a function of $\theta$.}}
\label{HoBT1}
\end{figure}

{Let us now consider the following initial state of the walker
$ \ket{\Psi_{1}(0)}=\frac{1}{\sqrt{2}}\left(\ket{0_{p}0_{c}}+\ket{1_{p}1_{c}}\right)$. This corresponds 
to an initially delocalized walker, over the site $x=0$ and $x=1$ in the position space, and entangled 
with the coin. Substituting $ \ket{\Psi_{1}(0)} $ in Eq.~(\ref{initial-k}) yields
\begin{equation}\label{entangledInitialk1}
\ket{\varphi ' _{k}(0)}=\frac{1}{\sqrt{2}}\left(\ket{0_{c}}+\re ^{-ik}\ket{1_{c}} \right),
\end{equation}
and 
\begin{align}
\cket{\varrho ' _{0}} =\ket{\varphi ' _{k}(0)}\bra{\varphi ' _{k}(0)}=\left(1, \cos k, -\sin k, 0 \right) ^{T}.
\end{align}
By exploiting Eq.~(\ref{pure-qfim-theorem}) one can evaluate the QFIm as follows 
\begin{align}\label{qfi-full-00}
\qfi_1=4 t^{2}\left(\begin{array}{cc}
\frac{\sin\theta}{1+\sin\theta} & 0\\ 
0 &1-\sin\theta\\
\end{array}\right),
\end{align}
where the initial-state dependent terms of Eq.~(\ref{pure-qfim-theorem}) vanish
\begin{align}
&\int _{-\pi}^{\pi}\frac{\rd k}{2\pi}\, \cbra{O_{\theta}}\mathcal{A}_{k}^{\mathbb{1}}\cket{\varrho ' _{0}}=\nonumber\\
&\int _{-\pi}^{\pi}{\frac{\rd k}{2\pi}\, \frac{2i\cos (k-\alpha)\sin (2k-\beta)\sin ^{2}\theta}{1-\cos ^{2}(k-\alpha)\cos ^{2}\theta}}=0,
\end{align}
and
\begin{align}
&\int _{-\pi}^{\pi}\frac{\rd k}{2\pi}\, \cbra{O_{\alpha}}\mathcal{A}_{k}^{\mathbb{1}}\cket{\varrho ' _{0}}=\nonumber\\
&\int _{-\pi}^{\pi}{\frac{\rd k}{2\pi}\, \frac{-i\sin (k-\alpha)\sin (2k-\beta)\sin 2\theta}{1-\cos ^{2}(k-\alpha)\cos ^{2}\theta}}=0.
\end{align}
In other words, having an initially entangled state makes the off-diagonal elements of the $ \qfi $ and the $
\ulc$ to vanish, and maximizes its diagonal elements (see Theorems~\ref{theorem1} and~\ref{theorem2}). 
The model is thus compatible and the inequalities \eref{eq:symmHolevo} are saturated, so that
\begin{align}
\rC^{H}(\Theta ,W)&=\rC^{S}(\Theta ,W)\nonumber\\
&=\Tr{\qfi_1^{-1}W}.
\end{align}
In particular, by assuming $ W= \mathbb{1}$, $ \Tr{\qfi_1^{-1}W} $ gives the sum of the mean square errors for each of the unknown parameters. One can thus calculate the Holevo bound 
{
\begin{align}\label{holevo-full-00}
\rC_1^{H}(\Theta , W)&=\frac{1}{t^{2}}g({\theta})=\frac{1}{t^{2}}\frac{\sin\theta+\cos ^{2}\theta}{4\sin\theta(1-\sin\theta)}.
\end{align}
}
Figure (\ref{HoBT1}) shows the Holevo bound for the full state for different values of $ \theta $. One will asymptotically gain the factor $ \frac{1}{t^{2}} $ which indicates a quadratic enhancement in precision.}

As another example, we consider the following initial state:
\begin{align}\label{ISexam2}
\ket{\Psi _{2}(0)}\hspace{-2pt}\bra{\Psi _{2}(0)}&=\ket{0}\hspace{-2pt}\bra{0}\otimes\ket{\gamma}\hspace{-2pt}\bra{\gamma}\\
\ket{\gamma}\hspace{-2pt}\bra{\gamma}&=\frac{1}{2}( \mathbb{1}+\cos\gamma\sigma _{x}+\sin\gamma\sigma _{y}). \nonumber
\end{align}
The localized initial condition of the walker sets us once again under the conditions for Corollary 4 to apply. This time, however, the initial state of the coin is parametrized by $\gamma$. The elements of the QFIm of the full space can be derived by means of (\ref{qfi-tt-full}-\ref{qfi-ta-full}) and are
\begin{align}
\qfi _{2,\theta\theta}&=\frac{4t^{2}}{1+\sin\theta}\left[\sin\theta -\frac{\sin ^{2}\theta}{1+\sin\theta}\sin ^{2}(\gamma-\phi)\right],\label{qfi-tt-full-e2}\\
\qfi _{2,\alpha\alpha}&= 4t^{2}\left(1-\sin\theta\right) \left[1-\frac{\sin ^{2}\theta}{1+\sin\theta}\cos ^{2}(\gamma-\phi)\right],\label{qfi-aa-full-e2}\\
\qfi_{2,\theta\alpha}&=\frac{-4t^{2}\left(1-\sin\theta\right)}{\cos\theta\left( 1+\sin\theta\right)}\left[\sin ^{2}\theta\sin(\gamma-\phi)\cos(\gamma-\phi)\right].\label{qfi-ta-full-e2}
\end{align}
{Note that by tunning $ \gamma $ in the initial state $ \ket{\Psi _{2}(0)} $ (Eq.~(\ref{ISexam2})), 
one finds the optimal value $ \gamma=\phi $ which maximizes $ \qfi _{2,\theta\theta} $ and makes 
the off-diagonal terms to vanish. On the other hand $ \qfi _{2,\alpha\alpha} $ is not maximum. Hence 
for this local initial state the compatibility conditions are not fulfilled}. 
\subsubsection{Joint estimation of two components of the magnetic field}
{We apply our formalism to estimate the components of a magnetic field \cite{Baum2016,Troiani2018,Apellaniz2018}. In order to do that, we consider a 
quantum walk in which the coin operation is governed by the magnetic field having 
two unknown components $ \vec{B}=(0,b_{2},b_{3}), $ where $ B=\sqrt{b_{2}^{2}+b_{3}^{2}} $. 
The magnetic field act therefore on a two-level system according to the following Hamiltonian 
\begin{equation}\label{magneticfield}
\re ^{-iB\hat{B}.\sigma}=\cos B \mathbb{1}-i\frac{\sin B}{B}b_{2}\sigma _{y}-i\frac{\sin B}{B}b_{3}\sigma _{z}.
\end{equation}
To estimate the unknown components of the magnetic field within our formalism, we encode them on the coin operator of the DTQW (Eq. (\ref{CoinOperator})) as follows 
\begin{align}\label{magnetic-coin}
\sin\theta &=-\frac{\sin B}{B}b_{2},\nonumber\\
\tan\alpha &=-\frac{\tan B}{B}b_{3},\nonumber\\
\beta &=0.
\end{align}
This implies that estimating the unknown parameters of the coin provides the components 
of the magnetic field.}
\subsubsection{Joint estimation of mass and charge in the Dirac equation}
{Our formalism may be used to simultaneous estimate physical quantities like mass and charge. 
In order to show this possibility, we consider the Dirac Hamiltonian in (1+1) dimensions
in the presence of an electromagnetic field (in Planck units $ \hbar=c=1 $)
\begin{equation}\label{DiracEq}
\left[i\gamma ^{\mu}\left( \partial _{\mu}-iqA_{\mu}\right)-m \right] \psi =0,
\end{equation}
where $ q $ and $ m $ denote the charge and the mass of a spinless particle, $ A_{\mu} $ is the vector potential, and the $ \gamma ^{\mu} $ denote Dirac gamma matrices with $ \mu=0,1 $, satisfying the 
anti-commutation relation $ \left\lbrace \gamma ^{\mu},\gamma ^{\nu}\right\rbrace=2g^{\mu\nu} \mathbb{1} $, 
in which $ g^{\mu\nu}=\text{diag}(1,-1) $ and $ \mathbb{1} $ is the $ 2\times 2 $ identity matrix. 
We choose $ \gamma ^{0}=\sigma _{x} $, $ \gamma ^{1}=-i\sigma _{y} $, and $ A^{\mu}=\left( 0, A_{x}\right)^{T} $, i.e. we assume a zero scalar potential. With the above assumptions, the Dirac Hamiltonian, Eq.~(\ref{DiracEq}), rewrites as 
\begin{equation}\label{DiracHam}
i\partial_{t}\psi =H_{D}\psi=\left( - i\sigma _{z}\partial _{x}+qA_{x}\sigma _{z}+m\sigma _{x}\right) \psi .
\end{equation}
The unitary evolution of the Dirac Hamiltonian for small $ \epsilon $ is given by \cite{Arrighi2014}
\begin{align}\label{DiracUni}
\ket{\psi (t+\epsilon)}&=\re ^{-iH_{D}\epsilon}\ket{\psi (t)}\nonumber\\
&=\re ^{-\epsilon\sigma _{z}\partial _{x}}\re ^{-i\epsilon (qA_{x}\sigma _{z}+m\sigma _{x})}\ket{\psi (t)}+O(\epsilon ^{2}),
\end{align}
where in the last line we have employed the Lie-Trotter product formula. Equation~(\ref{DiracUni}) shows that the evolution induced by the Dirac Hamiltonian corresponds to a DTQW where the first exponential term is the translational operator and the second one denotes the coin operator. Moreover, it indicates that the mass and charge of the particle corresponds to the coin parameters as follows
\begin{align}\label{dirac-coin}
\sin\theta &=\frac{-m}{\sqrt{q^{2}A_{x}^{2}+m^{2}}}\sin\left( \epsilon\sqrt{q^{2}A_{x}^{2}+m^{2}}\right)
\nonumber\\
&\simeq -m\epsilon,\nonumber\\
\tan\alpha &=\frac{-qA_{x}}{\sqrt{q^{2}A_{x}^{2}+m^{2}}}\tan\left( \epsilon\sqrt{q^{2}A_{x}^{2}+m^{2}}\right)
\nonumber\\
&\simeq -qA_{x}\epsilon,\nonumber\\
\beta &=\frac{\pi}{2}.
\end{align} 
DTQW thus represents a convenient tool to simulate the evolution of a Dirac particle, and to simultaneously estimate the mass and the charge of the particle via coin parameter estimation. Moreover, the vanishing of the Uhlmann curvature assures that the mass and the charge can be estimated simultaneously with fewer resources.}
\section{Conclusions}\label{Conclusion}
In this paper, we have addressed multi-parameter quantum estimation for one-dimensional discrete-time quantum walks. In particular, we have explored the possibility of exploiting  quantum walk to address the full statistical model, namely the simultaneous estimation of different unknown parameters. We have found the analytic expression of the quantum Fisher information matrix for the most general coin operator, and then exploited our findings to demonstrate that a (discrete-time)  quantum walker can be used as an optimal probe if the unknown parameters of the statistical model are encoded in the coin space. 
\par
We have shown that for the full model, only two out of the three parameters defining the coin operator can be actually  accessed, and proved that the resulting  two-parameter coin model is asymptotically classical i.e. the Uhlmann curvature vanishes. Finally, we have applied our findings to relevant case studies, including the simultaneous estimation of two components of a magnetic field and of the charge and the mass of a particle in the discretized Dirac model. Our results clarify the role of coin parameters in discrete quantum walks, and pave the way for further investigation in systems with more than a walker. 
\acknowledgments
M. G. A. P. is member of INdAM-GNFM.
\vfill \pagebreak \onecolumngrid
\appendix
\section{Proof of Eq. (\ref{1deriv-coinpart})}\label{appendix-1deriv-coinpart}
\noindent The first derivative of the coin part $ \left( \ket{\varphi _{k}^{t}(\Theta)}\right)  $ is given by
\begin{align}
\ket{\partial _{\mu}\varphi _{k}^{t}(\Theta)}&=\partial _{\mu} {u}_{k}^{t}\ket{\chi}_{c}\nonumber
=\sum _{m=0}^{t-1} {u}_{k}^{m}\, \mathbb{1}\,(\partial _{\mu} {u}_{k})\, {u}_{k}^{t-m-1}\ket{\chi}_{c}\nonumber
=\sum _{m=0}^{t-1} {u}_{k}^{m}\, {u}_{k} {u}_{k}^{\,\dagger}\,(\partial _{\mu} {u}_{k})\, {u}_{k}^{\dagger m+1} {u}_{k}^{t}\ket{\chi}_{c}\nonumber\\
&=\sum _{m=0}^{t-1} {u}_{k}^{m+1}\,\left(  {u}_{k}^{\,\dagger}\,\partial _{\mu} {u}_{k}\right) \, {u}_{k}^{\dagger m+1}\ket{\varphi _{k}^{t}(\Theta)},
\end{align}
\section{Useful relations in the calculation of the Bloch representation}\label{Useful-app}
\noindent Let us consider
\begin{align*}
A=\frac{1}{2}\left(a_0 \mathbb{1}+\vec{a}.\vec{\sigma}\right),\quad
B=\frac{1}{2}\left(b_0 \mathbb{1}+\vec{b}.\vec{\sigma}\right),\quad
C=\frac{1}{2}\left(c_0 \mathbb{1}+\vec{c}.\vec{\sigma}\right).
\end{align*}
It is it straightforward to show that
\begin{align}
\Tr{AB}&=\frac{1}{2}\cbraket{A^{\dagger}}{B}=\frac{1}{2}\left(a_{0}b_{0}+\vec{a}.\vec{b}\right),\label{TrAB}\\
\Tr{ABC}&=\frac{1}{4}\left(i\vec{a}.(\vec{b}\times\vec{c})+a_0\cbraket{B^{\dagger}}{C}+b_{0}\cbraket{C^{\dagger}}{A}+c_{0}\cbraket{A^{\dagger}}{B}-2 a_{0} b_{0} c_{0}\right)\nonumber\\
&=\frac{1}{4}\left(i\vec{a}.(\vec{b}\times\vec{c})+a_{0}(\vec{b}.\vec{c})+b_{0}(\vec{a}.\vec{c})+c_{0}(\vec{a}.\vec{b})+a_{0} b_{0} c_{0}\right)\label{TrABC}.
\end{align}
Using Eq. (\ref{TrABC}), we also have 
\begin{align}
\Tr{A\lbrace B,C\rbrace}&~~=\Tr{ABC+ACB} \overset{(\ref{TrABC})}{=}\frac{1}{2}\left[a_{0}(\vec{b}.\vec{c})+b_{0}(\vec{a}.\vec{c})+c_{0}(\vec{a}.\vec{b})+a_{0} b_{0} c_{0}\right],\label{Tr-iden-1}\\
\Tr{A [B,C]}&~~=\Tr{ABC-ACB} \overset{(\ref{TrABC})}{=}\frac{1}{2}i\vec{a}.\left(\vec{b}\times\vec{c}\right).\label{Tr-iden-2}
\end{align}
\section{Proof of theorem \ref{theorem1}}\label{theorem1Proof}
Let us define the following transformatio
\begin{align}
O'= {u}_{k}O {u}_{k}^{\dagger},\label{qoperation}
\equiv\mathcal{A}_{k}{O},
\end{align}
where $\mathcal{A}_{k} $ denotes the superoperator. From  
Eqs. (\ref{Uk}), (\ref{O-pauli}), and (\ref{4DVRofO}), one obtains 
\begin{align}
O'&= {u}_{k}O {u}_{k}^{\dagger}=\frac{1}{2}\left(
\begin{array}{cc}
 \re ^{-i (k-\alpha)}\cos\theta & \re ^{-i (k-\beta)}\sin\theta \\
 -\re ^{i (k-\beta)}\sin\theta & \re ^{i (k-\alpha)}\cos\theta\\
\end{array}
\right)\left(
\begin{array}{cc}
 o_0+ o_{z} & o_{x}-i o_{y}\\
o_{x}+io_{y}&o_0-o_{z}\\
\end{array}
\right)\left(
\begin{array}{cc}
\re ^{i (k-\alpha)}\cos\theta & -\re ^{-i (k-\beta)}\sin\theta \\
\re ^{i (k-\beta)}\sin\theta  & \re ^{-i (k-\alpha)}\cos\theta\\
\end{array}
\right),
\end{align}\label{Oprim-mat}
with four-dimensional column vector representation as
\begin{equation}\label{Oprim-vecform}
\cket{{O}'}=\left(\begin{array}{cccc}
o_0\\ 
\cos ^2(\theta ) (o_{x} \cos 2 (k-\alpha)-o_{y} \sin 2 (k-\alpha ))-\sin ^2\theta (o_{x} \cos 2 (k-\beta)+o_{y} \sin 2 (k-\beta ))-o_{z} \sin 2 \theta  \cos (2k-\alpha -\beta)\\
\cos ^2\theta (o_{x} \sin 2 (k-\alpha)+o_{y} \cos 2 (k-\alpha ))+\sin ^2\theta (o_{y} \cos 2 (k-\beta)-o_{x} \sin 2 (k-\beta ))-o_{z} \sin 2 \theta \sin (2k-\alpha -\beta)\\
\sin 2 \theta (o_{x} \cos (\alpha -\beta )+o_{y} \sin (\alpha -\beta ))+o_{z}\cos 2 \theta\\
\end{array}\right).
\end{equation}
We define $\cket{O'}=\mathcal{\tilde{A}}_{k}\cket{O}$ in which
\begin{align}\label{Matrix4-L}
\mathcal{\tilde{A}}_{k}=\left(\begin{array}{cccc}
1&0&0&0\\ 
0&\cos 2(k-\alpha)\,\cos ^{2}\theta -\cos 2(k-\beta)\,\sin ^{2}\theta &-\sin 2(k-\alpha)\,\cos ^{2}\theta-\sin 2(k-\beta)\,\sin ^{2}\theta & -\cos (2k-\alpha-\beta)\,\sin 2\theta\\
0&\sin 2(k-\alpha)\,\cos ^{2}\theta -\sin 2(k-\beta)\,\sin ^{2}\theta &\cos 2(k-\alpha)\,\cos ^{2}\theta +\cos 2(k-\beta)\,\sin ^{2}\theta & -\sin (2k-\alpha-\beta)\,\sin 2\theta\\
0&\cos (\alpha -\beta)\,\sin 2\theta &\sin (\alpha -\beta)\,\sin 2\theta &\cos 2\theta\\
\end{array}\right).
\end{align}
In order to evaluate Eq. (\ref{pure-qrw-fund}), we need to calculate $\mathcal{A}_{k}^{'}=\sum_{\mu =0}^{t-1}\mathcal{A}_{k}^{\mu +1}$ and $\mathcal{A}_{k}^t$ . The spectral decomposition of $\mathcal{\tilde{A}}_{k}$ yields
\begin{equation}\label{Spectral-L}
\mathcal{\tilde{A}}_{k}=\ket{\lambda _{1}}\bra{\lambda _{1}}+\ket{\lambda _{2}}\bra{\lambda _{2}}+e^{2i\omega}\ket{\lambda _{3}}\bra{\lambda _{3}}+
e^{-2i\omega}\ket{\lambda _{4}}\bra{\lambda _{4}},
\end{equation}
where the eigenvalues of $ \mathcal{\tilde{A}}_{k} $ are
\begin{align}
\lambda _{1}=\lambda _{2}=1,~~~~~~\lambda _{3}=e^{2i\omega},~~~~~~\lambda _{4}=e^{-2i\omega},
\end{align}
in which $ \cos\omega\equiv\cos (k-\alpha)\cos\theta $. From whence
\begin{align}\label{superop-deriv-SDecom-1}
\mathcal{\tilde{A}}_{k}^{'}=\sum_{\mu =0}^{t-1}\mathcal{\tilde{A}}_{k}^{\mu +1}&=t\left(\ket{\lambda _{1}}\bra{\lambda _{1}}+\ket{\lambda _{2}}\bra{\lambda _{2}}\right)+
\sum_{\mu =0}^{t-1}e^{2i\omega(\mu +1)}\ket{\lambda _{3}}\bra{\lambda _{3}}+\sum_{\mu =0}^{t-1}e^{-2i\omega(\mu +1)}\ket{\lambda _{4}}\bra{\lambda _{4}}\nonumber\\
&=t\left(\ket{\lambda _{1}}\bra{\lambda _{1}}+\ket{\lambda _{2}}\bra{\lambda _{2}}\right)+
\frac{e^{2i\omega}\left(1-e^{2i\omega t}\right) }{1-e^{2i\omega}}\ket{\lambda _{3}}\bra{\lambda _{3}}+\frac{e^{-2i\omega}\left(1-e^{-2i\omega t}\right) }{1-e^{-2i\omega}}\ket{\lambda _{4}}\bra{\lambda _{4}},
\end{align}
and
\begin{equation}\label{Spectral-Lpowt}
\mathcal{\tilde{A}}_{k}^{t}=\ket{\lambda _{1}}\bra{\lambda _{1}}+\ket{\lambda _{2}}\bra{\lambda _{2}}+e^{2i\omega t}\ket{\lambda _{3}}\bra{\lambda _{3}}+
e^{-2i\omega t}\ket{\lambda _{4}}\bra{\lambda _{4}}.
\end{equation}
Substituting Eqs. (\ref{superop-deriv-SDecom-1}) and (\ref{Spectral-Lpowt}) in Eq. (\ref{pure-qrw-fund}) and neglecting fast oscillation terms (proportional to $e^{\pm2i\omega t}$) of integrals in the asymptotic limit ($ t\gg 1 $), reveals this fact that only the eigenvecotors corresponding to the $ \lambda _{1} $ and $ \lambda _{2} $ should be considered. Hence in the asymptotic limit
\begin{align}
\mathcal{\tilde{A}}_{k}^{'}&=t\mathcal{\tilde{A}}_{k}^{\mathbb{1}},\\
\mathcal{\tilde{A}}_{k}^{t}&=\mathcal{\tilde{A}}_{k}^{\mathbb{1}},
\end{align}
where $\mathcal{\tilde{A}}_{k}^{\mathbb{1}}=\ket{\lambda _{1}}\bra{\lambda _{1}}+\ket{\lambda _{2}}\bra{\lambda _{2}}$ is a projector of $\mathcal{\tilde{A}}_{k}$ in the subspace with the eigenvalue of 1 and
\begin{align}
\ket{\lambda _{1}}&=\frac{1}{\sqrt{2(1-\cos\omega)}}\left(\begin{array}{cccc}
\cos (k-\alpha)\,\cos\theta -1\\ 
\sin (k-\beta)\,\sin\theta\\
-\cos (k-\beta)\,\sin\theta\\
\sin (k-\alpha)\,\cos\theta\\
\end{array}\right),\label{EV-L1}\\
\ket{\lambda _{2}}&=\frac{1}{\sqrt{2(1+\cos \omega)}}\left(\begin{array}{cccc}
\cos (k-\alpha)\,\cos\theta +1\\ 
\sin (k-\beta)\,\sin\theta\\
-\cos (k-\beta)\,\sin\theta\\
\sin (k-\alpha)\,\cos\theta\\
\end{array}\right).\label{EV-L2}
\end{align} 
Applying these relations to Eq. (\ref{pure-qrw-fund}) yields 
\begin{align}\label{pure-qfim-4V}
\frac{t^2}{2\pi}\int _{-\pi}^{\pi}\rd k\, \Tr{{\mathcal{A}_{k}^{\mathbb{1}}{\left(O_{\mu}^{\dagger}\right)}}\mathcal{A}_{k}^{\mathbb{1}}{\left(O_{\nu}\right)}{\mathcal{A}_{k}^{\mathbb{1}}}{\left( \varrho_0\right)}}-\left(\frac{t}{2\pi}\int _{-\pi}^{\pi}\rd k\, \Tr{{\mathcal{A}_{k}^{\mathbb{1}}{\left(O_{\mu}^{\dagger}\right)}}{\mathcal{A}_{k}^{\mathbb{1}}}{\left( \varrho_0\right)}} \right)\left(\frac{t}{2\pi}\int _{-\pi}^{\pi}\rd k\, \Tr{{\mathcal{A}_{k}^{\mathbb{1}}{\left(O_{\nu}\right)}}{\mathcal{A}_{k}^{\mathbb{1}}}{\left( \varrho_0\right)}} \right).
\end{align}
By defining ${\mathcal{A}_{k}^{\mathbb{1}}{\left(O_{\mu}^{\dagger}\right)}}\equiv A,\,\mathcal{A}_{k}^{\mathbb{1}}{\left(O_{\nu}\right)}\equiv B$, ${\mathcal{A}_{k}^{\mathbb{1}}}{\left( \varrho_0\right)}\equiv C$, 
and noting Eq. (\ref{TrABC}), we have
\begin{align}
\Tr{{\mathcal{A}_{k}^{\mathbb{1}}{\left(O_{\mu}^{\dagger}\right)}}\mathcal{A}_{k}^{\mathbb{1}}{\left(O_{\nu}\right)}{\mathcal{A}_{k}^{\mathbb{1}}}{\left( \varrho_0\right)}}&=\frac{1}{4}\cbra{O_\mu}\mathcal{\tilde{A}}_{k}^{\mathbb{1}}\cket{O_\nu},\nonumber\\
\Tr{{\mathcal{A}_{k}^{\mathbb{1}}{\left(O_{\mu}^{\dagger}\right)}}{\mathcal{A}_{k}^{\mathbb{1}}}{\left( \varrho_0\right)}}&=\frac{1}{2}\cbra{O_\mu}\mathcal{\tilde{A}}_{k}^{\mathbb{1}}\cket{ \varrho_0},\nonumber\\
\Tr{{\mathcal{A}_{k}^{\mathbb{1}}{\left(O_{\nu}\right)}}{\mathcal{A}_{k}^{\mathbb{1}}}{\left( \varrho_0\right)}}&=\frac{1}{2}\cbra{ \varrho_0}\mathcal{\tilde{A}}_{k}^{\mathbb{1}}\cket{O_\nu},
\end{align}
in which we use the facts that $a_0=b_0=0, c_0=1, \left({\mathcal{\tilde{A}}_{k}^{\mathbb{1}}}\right)^2=\mathcal{\tilde{A}}_{k}^{\mathbb{1}}$, and
\begin{align}
\vec{a}.(\vec{b}\times\vec{c})=\mathrm{Det}\left(\mathcal{\tilde{A}}_{k}^{\mathbb{1}}\right)\left(\vec{{O_\mu}}.(\vec{{O_\nu}}\times\vec{{ \varrho_0}})\right)=0.
\end{align}
Noting that the explicit form of $\mathcal{\tilde{A}}_{k}^{\mathbb{1}}$ implies that $\mathrm{Det}\left(\mathcal{A}_{k}^{\mathbb{1}}\right)=0$. Hence the simple form of Eq. (\ref{pure-qrw-fund}) is
\begin{align}\label{pure-qfim-simplified}
\average{\partial _{\mu}\Psi\Big\vert\partial _{\nu}\Psi}-\average{\partial _{\mu}\Psi\Big\vert\Psi}\average{\Psi\Big\vert\partial _{\nu}\Psi}=\frac{t^2}{4}\left\lbrace\int _{-\pi}^{\pi}\frac{\rd k}{2\pi}\, \cbra{O_\mu}\mathcal{\tilde{A}}_{k}^{\mathbb{1}}\cket{O_\nu}-\left(\int _{-\pi}^{\pi}\frac{\rd k}{2\pi}\, \cbra{O_\mu}\mathcal{\tilde{A}}_{k}^{\mathbb{1}}\cket{ \varrho_0} \right)\left(\int _{-\pi}^{\pi}\frac{\rd k}{2\pi}\, \cbra{ \varrho_0}\mathcal{\tilde{A}}_{k}^{\mathbb{1}}\cket{O_\nu} \right)\right\rbrace.
\end{align}
We remark that $O_i$ is the anti-Hermitian operator and the elements of $\cket{O_i}$ are purely imaginary. Consequently Eq. (\ref{pure-qfim-simplified}) will be a real expression. Then $\ulc _{\mu\nu}=0$---see Eq. (\ref{ulc-pure}).

\bibliography{Ref}

\begin{thebibliography}{63}%
\makeatletter
\providecommand \@ifxundefined [1]{%
 \@ifx{#1\undefined}
}%
\providecommand \@ifnum [1]{%
 \ifnum #1\expandafter \@firstoftwo
 \else \expandafter \@secondoftwo
 \fi
}%
\providecommand \@ifx [1]{%
 \ifx #1\expandafter \@firstoftwo
 \else \expandafter \@secondoftwo
 \fi
}%
\providecommand \natexlab [1]{#1}%
\providecommand \enquote  [1]{``#1''}%
\providecommand \bibnamefont  [1]{#1}%
\providecommand \bibfnamefont [1]{#1}%
\providecommand \citenamefont [1]{#1}%
\providecommand \href@noop [0]{\@secondoftwo}%
\providecommand \href [0]{\begingroup \@sanitize@url \@href}%
\providecommand \@href[1]{\@@startlink{#1}\@@href}%
\providecommand \@@href[1]{\endgroup#1\@@endlink}%
\providecommand \@sanitize@url [0]{\catcode `\\12\catcode `\$12\catcode
  `\&12\catcode `\#12\catcode `\^12\catcode `\_12\catcode `\%12\relax}%
\providecommand \@@startlink[1]{}%
\providecommand \@@endlink[0]{}%
\providecommand \url  [0]{\begingroup\@sanitize@url \@url }%
\providecommand \@url [1]{\endgroup\@href {#1}{\urlprefix }}%
\providecommand \urlprefix  [0]{URL }%
\providecommand \Eprint [0]{\href }%
\providecommand \doibase [0]{http://dx.doi.org/}%
\providecommand \selectlanguage [0]{\@gobble}%
\providecommand \bibinfo  [0]{\@secondoftwo}%
\providecommand \bibfield  [0]{\@secondoftwo}%
\providecommand \translation [1]{[#1]}%
\providecommand \BibitemOpen [0]{}%
\providecommand \bibitemStop [0]{}%
\providecommand \bibitemNoStop [0]{.\EOS\space}%
\providecommand \EOS [0]{\spacefactor3000\relax}%
\providecommand \BibitemShut  [1]{\csname bibitem#1\endcsname}%
\let\auto@bib@innerbib\@empty
\bibitem [{\citenamefont {Giovannetti}\ \emph {et~al.}(2011)\citenamefont
  {Giovannetti}, \citenamefont {Lloyd},\ and\ \citenamefont
  {Maccone}}]{Giovannetti2011advances}%
  \BibitemOpen
  \bibfield  {author} {\bibinfo {author} {\bibfnamefont {V.}~\bibnamefont
  {Giovannetti}}, \bibinfo {author} {\bibfnamefont {S.}~\bibnamefont {Lloyd}},
  \ and\ \bibinfo {author} {\bibfnamefont {L.}~\bibnamefont {Maccone}},\ }\href
  {\doibase https://doi.org/10.1038/nphoton.2011.35} {\bibfield  {journal}
  {\bibinfo  {journal} {Nature photonics}\ }\textbf {\bibinfo {volume} {5}},\
  \bibinfo {pages} {222} (\bibinfo {year} {2011})}\BibitemShut {NoStop}%
\bibitem [{\citenamefont {Paris}(2009)}]{Paris2009quantum}%
  \BibitemOpen
  \bibfield  {author} {\bibinfo {author} {\bibfnamefont {M.~G.~A.}\
  \bibnamefont {Paris}},\ }\href {\doibase 10.1142/S0219749909004839}
  {\bibfield  {journal} {\bibinfo  {journal} {International Journal of Quantum
  Information}\ }\textbf {\bibinfo {volume} {07}},\ \bibinfo {pages} {125}
  (\bibinfo {year} {2009})}\BibitemShut {NoStop}%
\bibitem [{\citenamefont {Degen}\ \emph {et~al.}(2017)\citenamefont {Degen},
  \citenamefont {Reinhard},\ and\ \citenamefont {Cappellaro}}]{Degen2017}%
  \BibitemOpen
  \bibfield  {author} {\bibinfo {author} {\bibfnamefont {C.~L.}\ \bibnamefont
  {Degen}}, \bibinfo {author} {\bibfnamefont {F.}~\bibnamefont {Reinhard}}, \
  and\ \bibinfo {author} {\bibfnamefont {P.}~\bibnamefont {Cappellaro}},\
  }\href {\doibase 10.1103/RevModPhys.89.035002} {\bibfield  {journal}
  {\bibinfo  {journal} {Rev. Mod. Phys.}\ }\textbf {\bibinfo {volume} {89}},\
  \bibinfo {pages} {035002} (\bibinfo {year} {2017})}\BibitemShut {NoStop}%
\bibitem [{\citenamefont {Escher}\ \emph {et~al.}(2011)\citenamefont {Escher},
  \citenamefont {de~Matos~Filho},\ and\ \citenamefont
  {Davidovich}}]{Escher2011}%
  \BibitemOpen
  \bibfield  {author} {\bibinfo {author} {\bibfnamefont {B.}~\bibnamefont
  {Escher}}, \bibinfo {author} {\bibfnamefont {R.}~\bibnamefont
  {de~Matos~Filho}}, \ and\ \bibinfo {author} {\bibfnamefont {L.}~\bibnamefont
  {Davidovich}},\ }\href {\doibase https://doi.org/10.1038/nphys1958}
  {\bibfield  {journal} {\bibinfo  {journal} {Nature Physics}\ }\textbf
  {\bibinfo {volume} {7}},\ \bibinfo {pages} {406} (\bibinfo {year}
  {2011})}\BibitemShut {NoStop}%
\bibitem [{\citenamefont {Braun}\ \emph {et~al.}(2018)\citenamefont {Braun},
  \citenamefont {Adesso}, \citenamefont {Benatti}, \citenamefont {Floreanini},
  \citenamefont {Marzolino}, \citenamefont {Mitchell},\ and\ \citenamefont
  {Pirandola}}]{Braun2018}%
  \BibitemOpen
  \bibfield  {author} {\bibinfo {author} {\bibfnamefont {D.}~\bibnamefont
  {Braun}}, \bibinfo {author} {\bibfnamefont {G.}~\bibnamefont {Adesso}},
  \bibinfo {author} {\bibfnamefont {F.}~\bibnamefont {Benatti}}, \bibinfo
  {author} {\bibfnamefont {R.}~\bibnamefont {Floreanini}}, \bibinfo {author}
  {\bibfnamefont {U.}~\bibnamefont {Marzolino}}, \bibinfo {author}
  {\bibfnamefont {M.~W.}\ \bibnamefont {Mitchell}}, \ and\ \bibinfo {author}
  {\bibfnamefont {S.}~\bibnamefont {Pirandola}},\ }\href {\doibase
  10.1103/RevModPhys.90.035006} {\bibfield  {journal} {\bibinfo  {journal}
  {Rev. Mod. Phys.}\ }\textbf {\bibinfo {volume} {90}},\ \bibinfo {pages}
  {035006} (\bibinfo {year} {2018})}\BibitemShut {NoStop}%
\bibitem [{\citenamefont {Caves}(1981)}]{Caves1981}%
  \BibitemOpen
  \bibfield  {author} {\bibinfo {author} {\bibfnamefont {C.~M.}\ \bibnamefont
  {Caves}},\ }\href {\doibase 10.1103/PhysRevD.23.1693} {\bibfield  {journal}
  {\bibinfo  {journal} {Phys. Rev. D}\ }\textbf {\bibinfo {volume} {23}},\
  \bibinfo {pages} {1693} (\bibinfo {year} {1981})}\BibitemShut {NoStop}%
\bibitem [{\citenamefont {Demkowicz-Dobrza{\'{n}}ski}\ \emph
  {et~al.}(2015)\citenamefont {Demkowicz-Dobrza{\'{n}}ski}, \citenamefont
  {Jarzyna},\ and\ \citenamefont {Ko{\l}ody{\'{n}}ski}}]{RafalPO}%
  \BibitemOpen
  \bibfield  {author} {\bibinfo {author} {\bibfnamefont {R.}~\bibnamefont
  {Demkowicz-Dobrza{\'{n}}ski}}, \bibinfo {author} {\bibfnamefont
  {M.}~\bibnamefont {Jarzyna}}, \ and\ \bibinfo {author} {\bibfnamefont
  {J.}~\bibnamefont {Ko{\l}ody{\'{n}}ski}},\ }\href {\doibase
  10.1016/bs.po.2015.02.003} {\bibfield  {journal} {\bibinfo  {journal} {Prog.
  Opt.}\ }\textbf {\bibinfo {volume} {60}},\ \bibinfo {pages} {345} (\bibinfo
  {year} {2015})}\BibitemShut {NoStop}%
\bibitem [{\citenamefont {Acernese}\ \emph {et~al.}(2019)\citenamefont
  {Acernese} \emph {et~al.}}]{Acernese2019}%
  \BibitemOpen
  \bibfield  {author} {\bibinfo {author} {\bibfnamefont {F.}~\bibnamefont
  {Acernese}} \emph {et~al.},\ }\href {\doibase 10.1103/PhysRevLett.123.231108}
  {\bibfield  {journal} {\bibinfo  {journal} {Phys. Rev. Lett.}\ }\textbf
  {\bibinfo {volume} {123}},\ \bibinfo {pages} {231108} (\bibinfo {year}
  {2019})}\BibitemShut {NoStop}%
\bibitem [{\citenamefont {Tse}\ \emph {et~al.}(2019)\citenamefont {Tse} \emph
  {et~al.}}]{Tse2019}%
  \BibitemOpen
  \bibfield  {author} {\bibinfo {author} {\bibfnamefont {M.}~\bibnamefont
  {Tse}} \emph {et~al.},\ }\href {\doibase 10.1103/PhysRevLett.123.231107}
  {\bibfield  {journal} {\bibinfo  {journal} {Phys. Rev. Lett.}\ }\textbf
  {\bibinfo {volume} {123}},\ \bibinfo {pages} {231107} (\bibinfo {year}
  {2019})}\BibitemShut {NoStop}%
\bibitem [{\citenamefont {Taylor}\ and\ \citenamefont
  {Bowen}(2016)}]{Taylor2016}%
  \BibitemOpen
  \bibfield  {author} {\bibinfo {author} {\bibfnamefont {M.~A.}\ \bibnamefont
  {Taylor}}\ and\ \bibinfo {author} {\bibfnamefont {W.~P.}\ \bibnamefont
  {Bowen}},\ }\href {\doibase 10.1016/j.physrep.2015.12.002} {\bibfield
  {journal} {\bibinfo  {journal} {Phys. Rep.}\ }\textbf {\bibinfo {volume}
  {615}},\ \bibinfo {pages} {1} (\bibinfo {year} {2016})},\ \Eprint
  {http://arxiv.org/abs/1409.0950} {arXiv:1409.0950} \BibitemShut {NoStop}%
\bibitem [{\citenamefont {Budker}\ and\ \citenamefont
  {Romalis}(2007)}]{Budker2007}%
  \BibitemOpen
  \bibfield  {author} {\bibinfo {author} {\bibfnamefont {D.}~\bibnamefont
  {Budker}}\ and\ \bibinfo {author} {\bibfnamefont {M.}~\bibnamefont
  {Romalis}},\ }\href {\doibase 10.1038/nphys566} {\bibfield  {journal}
  {\bibinfo  {journal} {Nat. Phys.}\ }\textbf {\bibinfo {volume} {3}},\
  \bibinfo {pages} {227} (\bibinfo {year} {2007})},\ \Eprint
  {http://arxiv.org/abs/0611246} {arXiv:0611246 [physics]} \BibitemShut
  {NoStop}%
\bibitem [{\citenamefont {Koschorreck}\ \emph {et~al.}(2010)\citenamefont
  {Koschorreck}, \citenamefont {Napolitano}, \citenamefont {Dubost},\ and\
  \citenamefont {Mitchell}}]{Koschorreck2010}%
  \BibitemOpen
  \bibfield  {author} {\bibinfo {author} {\bibfnamefont {M.}~\bibnamefont
  {Koschorreck}}, \bibinfo {author} {\bibfnamefont {M.}~\bibnamefont
  {Napolitano}}, \bibinfo {author} {\bibfnamefont {B.}~\bibnamefont {Dubost}},
  \ and\ \bibinfo {author} {\bibfnamefont {M.~W.}\ \bibnamefont {Mitchell}},\
  }\href {\doibase 10.1103/PhysRevLett.104.093602} {\bibfield  {journal}
  {\bibinfo  {journal} {Phys. Rev. Lett.}\ }\textbf {\bibinfo {volume} {104}},\
  \bibinfo {pages} {093602} (\bibinfo {year} {2010})}\BibitemShut {NoStop}%
\bibitem [{\citenamefont {Wasilewski}\ \emph {et~al.}(2010)\citenamefont
  {Wasilewski}, \citenamefont {Jensen}, \citenamefont {Krauter}, \citenamefont
  {Renema}, \citenamefont {Balabas},\ and\ \citenamefont
  {Polzik}}]{Wasilewski2010}%
  \BibitemOpen
  \bibfield  {author} {\bibinfo {author} {\bibfnamefont {W.}~\bibnamefont
  {Wasilewski}}, \bibinfo {author} {\bibfnamefont {K.}~\bibnamefont {Jensen}},
  \bibinfo {author} {\bibfnamefont {H.}~\bibnamefont {Krauter}}, \bibinfo
  {author} {\bibfnamefont {J.~J.}\ \bibnamefont {Renema}}, \bibinfo {author}
  {\bibfnamefont {M.~V.}\ \bibnamefont {Balabas}}, \ and\ \bibinfo {author}
  {\bibfnamefont {E.~S.}\ \bibnamefont {Polzik}},\ }\href {\doibase
  10.1103/PhysRevLett.104.133601} {\bibfield  {journal} {\bibinfo  {journal}
  {Phys. Rev. Lett.}\ }\textbf {\bibinfo {volume} {104}},\ \bibinfo {pages}
  {133601} (\bibinfo {year} {2010})}\BibitemShut {NoStop}%
\bibitem [{\citenamefont {Sewell}\ \emph {et~al.}(2012)\citenamefont {Sewell},
  \citenamefont {Koschorreck}, \citenamefont {Napolitano}, \citenamefont
  {Dubost}, \citenamefont {Behbood},\ and\ \citenamefont
  {Mitchell}}]{Sewell2012}%
  \BibitemOpen
  \bibfield  {author} {\bibinfo {author} {\bibfnamefont {R.~J.}\ \bibnamefont
  {Sewell}}, \bibinfo {author} {\bibfnamefont {M.}~\bibnamefont {Koschorreck}},
  \bibinfo {author} {\bibfnamefont {M.}~\bibnamefont {Napolitano}}, \bibinfo
  {author} {\bibfnamefont {B.}~\bibnamefont {Dubost}}, \bibinfo {author}
  {\bibfnamefont {N.}~\bibnamefont {Behbood}}, \ and\ \bibinfo {author}
  {\bibfnamefont {M.~W.}\ \bibnamefont {Mitchell}},\ }\href {\doibase
  10.1103/PhysRevLett.109.253605} {\bibfield  {journal} {\bibinfo  {journal}
  {Phys. Rev. Lett.}\ }\textbf {\bibinfo {volume} {109}},\ \bibinfo {pages}
  {253605} (\bibinfo {year} {2012})},\ \Eprint {http://arxiv.org/abs/1111.6969}
  {1111.6969} \BibitemShut {NoStop}%
\bibitem [{\citenamefont {Troiani}\ and\ \citenamefont
  {Paris}(2018)}]{Troiani2018}%
  \BibitemOpen
  \bibfield  {author} {\bibinfo {author} {\bibfnamefont {F.}~\bibnamefont
  {Troiani}}\ and\ \bibinfo {author} {\bibfnamefont {M.~G.~A.}\ \bibnamefont
  {Paris}},\ }\href {\doibase 10.1103/PhysRevLett.120.260503} {\bibfield
  {journal} {\bibinfo  {journal} {Phys. Rev. Lett.}\ }\textbf {\bibinfo
  {volume} {120}},\ \bibinfo {pages} {260503} (\bibinfo {year}
  {2018})}\BibitemShut {NoStop}%
\bibitem [{\citenamefont {Ludlow}\ \emph {et~al.}(2015)\citenamefont {Ludlow},
  \citenamefont {Boyd}, \citenamefont {Ye}, \citenamefont {Peik},\ and\
  \citenamefont {Schmidt}}]{Ludlow2015}%
  \BibitemOpen
  \bibfield  {author} {\bibinfo {author} {\bibfnamefont {A.~D.}\ \bibnamefont
  {Ludlow}}, \bibinfo {author} {\bibfnamefont {M.~M.}\ \bibnamefont {Boyd}},
  \bibinfo {author} {\bibfnamefont {J.}~\bibnamefont {Ye}}, \bibinfo {author}
  {\bibfnamefont {E.}~\bibnamefont {Peik}}, \ and\ \bibinfo {author}
  {\bibfnamefont {P.~O.}\ \bibnamefont {Schmidt}},\ }\href {\doibase
  10.1103/RevModPhys.87.637} {\bibfield  {journal} {\bibinfo  {journal} {Rev.
  Mod. Phys.}\ }\textbf {\bibinfo {volume} {87}},\ \bibinfo {pages} {637}
  (\bibinfo {year} {2015})}\BibitemShut {NoStop}%
\bibitem [{\citenamefont {Louchet-Chauvet}\ \emph {et~al.}(2010)\citenamefont
  {Louchet-Chauvet}, \citenamefont {Appel}, \citenamefont {Renema},
  \citenamefont {Oblak}, \citenamefont {Kjaergaard},\ and\ \citenamefont
  {Polzik}}]{Louchet-Chauvet2010}%
  \BibitemOpen
  \bibfield  {author} {\bibinfo {author} {\bibfnamefont {A.}~\bibnamefont
  {Louchet-Chauvet}}, \bibinfo {author} {\bibfnamefont {J.}~\bibnamefont
  {Appel}}, \bibinfo {author} {\bibfnamefont {J.~J.}\ \bibnamefont {Renema}},
  \bibinfo {author} {\bibfnamefont {D.}~\bibnamefont {Oblak}}, \bibinfo
  {author} {\bibfnamefont {N.}~\bibnamefont {Kjaergaard}}, \ and\ \bibinfo
  {author} {\bibfnamefont {E.~S.}\ \bibnamefont {Polzik}},\ }\href {\doibase
  10.1088/1367-2630/12/6/065032} {\bibfield  {journal} {\bibinfo  {journal}
  {New J. Phys.}\ }\textbf {\bibinfo {volume} {12}},\ \bibinfo {pages} {065032}
  (\bibinfo {year} {2010})}\BibitemShut {NoStop}%
\bibitem [{\citenamefont {Kessler}\ \emph {et~al.}(2014)\citenamefont
  {Kessler}, \citenamefont {K{\'{o}}m{\'{a}}r}, \citenamefont {Bishof},
  \citenamefont {Jiang}, \citenamefont {S{\o}rensen}, \citenamefont {Ye},\ and\
  \citenamefont {Lukin}}]{Kessler2014a}%
  \BibitemOpen
  \bibfield  {author} {\bibinfo {author} {\bibfnamefont {E.~M.}\ \bibnamefont
  {Kessler}}, \bibinfo {author} {\bibfnamefont {P.}~\bibnamefont
  {K{\'{o}}m{\'{a}}r}}, \bibinfo {author} {\bibfnamefont {M.}~\bibnamefont
  {Bishof}}, \bibinfo {author} {\bibfnamefont {L.}~\bibnamefont {Jiang}},
  \bibinfo {author} {\bibfnamefont {A.~S.}\ \bibnamefont {S{\o}rensen}},
  \bibinfo {author} {\bibfnamefont {J.}~\bibnamefont {Ye}}, \ and\ \bibinfo
  {author} {\bibfnamefont {M.~D.}\ \bibnamefont {Lukin}},\ }\href {\doibase
  10.1103/PhysRevLett.112.190403} {\bibfield  {journal} {\bibinfo  {journal}
  {Phys. Rev. Lett.}\ }\textbf {\bibinfo {volume} {112}},\ \bibinfo {pages}
  {190403} (\bibinfo {year} {2014})},\ \Eprint {http://arxiv.org/abs/1310.6043}
  {arXiv:1310.6043} \BibitemShut {NoStop}%
\bibitem [{\citenamefont {Tamascelli}\ \emph {et~al.}(2020)\citenamefont
  {Tamascelli}, \citenamefont {Benedetti}, \citenamefont {Breuer},\ and\
  \citenamefont {Paris}}]{Tamascelli2020}%
  \BibitemOpen
  \bibfield  {author} {\bibinfo {author} {\bibfnamefont {D.}~\bibnamefont
  {Tamascelli}}, \bibinfo {author} {\bibfnamefont {C.}~\bibnamefont
  {Benedetti}}, \bibinfo {author} {\bibfnamefont {H.-P.}\ \bibnamefont
  {Breuer}}, \ and\ \bibinfo {author} {\bibfnamefont {M.~G.~A.}\ \bibnamefont
  {Paris}},\ }\href {\doibase 10.1088/1367-2630/aba0e5} {\bibfield  {journal}
  {\bibinfo  {journal} {New J. Phys.}\ }\textbf {\bibinfo {volume} {22}},\
  \bibinfo {pages} {083027} (\bibinfo {year} {2020})}\BibitemShut {NoStop}%
\bibitem [{\citenamefont {Tamascelli}\ \emph {et~al.}(2016)\citenamefont
  {Tamascelli}, \citenamefont {Benedetti}, \citenamefont {Olivares},\ and\
  \citenamefont {Paris}}]{tama16}%
  \BibitemOpen
  \bibfield  {author} {\bibinfo {author} {\bibfnamefont {D.}~\bibnamefont
  {Tamascelli}}, \bibinfo {author} {\bibfnamefont {C.}~\bibnamefont
  {Benedetti}}, \bibinfo {author} {\bibfnamefont {S.}~\bibnamefont {Olivares}},
  \ and\ \bibinfo {author} {\bibfnamefont {M.~G.~A.}\ \bibnamefont {Paris}},\
  }\href {\doibase 10.1103/PhysRevA.94.042129} {\bibfield  {journal} {\bibinfo
  {journal} {Phys. Rev. A}\ }\textbf {\bibinfo {volume} {94}},\ \bibinfo
  {pages} {042129} (\bibinfo {year} {2016})}\BibitemShut {NoStop}%
\bibitem [{\citenamefont {Giovannetti}\ \emph {et~al.}(2004)\citenamefont
  {Giovannetti}, \citenamefont {Lloyd},\ and\ \citenamefont
  {Maccone}}]{Giovannetti2004Sci}%
  \BibitemOpen
  \bibfield  {author} {\bibinfo {author} {\bibfnamefont {V.}~\bibnamefont
  {Giovannetti}}, \bibinfo {author} {\bibfnamefont {S.}~\bibnamefont {Lloyd}},
  \ and\ \bibinfo {author} {\bibfnamefont {L.}~\bibnamefont {Maccone}},\ }\href
  {\doibase 10.1126/science.1104149} {\bibfield  {journal} {\bibinfo  {journal}
  {Science}\ }\textbf {\bibinfo {volume} {306}},\ \bibinfo {pages} {1330}
  (\bibinfo {year} {2004})}\BibitemShut {NoStop}%
\bibitem [{\citenamefont {Giovannetti}\ \emph {et~al.}(2006)\citenamefont
  {Giovannetti}, \citenamefont {Lloyd},\ and\ \citenamefont
  {Maccone}}]{Giovannetti2006prl}%
  \BibitemOpen
  \bibfield  {author} {\bibinfo {author} {\bibfnamefont {V.}~\bibnamefont
  {Giovannetti}}, \bibinfo {author} {\bibfnamefont {S.}~\bibnamefont {Lloyd}},
  \ and\ \bibinfo {author} {\bibfnamefont {L.}~\bibnamefont {Maccone}},\ }\href
  {\doibase 10.1103/PhysRevLett.96.010401} {\bibfield  {journal} {\bibinfo
  {journal} {Phys. Rev. Lett.}\ }\textbf {\bibinfo {volume} {96}},\ \bibinfo
  {pages} {010401} (\bibinfo {year} {2006})}\BibitemShut {NoStop}%
\bibitem [{\citenamefont {Demkowicz-Dobrzański}\ \emph
  {et~al.}(2015)\citenamefont {Demkowicz-Dobrzański}, \citenamefont
  {Jarzyna},\ and\ \citenamefont {Kołodyński}}]{Demko2015}%
  \BibitemOpen
  \bibfield  {author} {\bibinfo {author} {\bibfnamefont {R.}~\bibnamefont
  {Demkowicz-Dobrzański}}, \bibinfo {author} {\bibfnamefont {M.}~\bibnamefont
  {Jarzyna}}, \ and\ \bibinfo {author} {\bibfnamefont {J.}~\bibnamefont
  {Kołodyński}},\ }\href {\doibase https://doi.org/10.1016/bs.po.2015.02.003}
  {\ \bibinfo {series} {Progress in Optics},\ \textbf {\bibinfo {volume}
  {60}},\ \bibinfo {pages} {345 } (\bibinfo {year} {2015})}\BibitemShut
  {NoStop}%
\bibitem [{\citenamefont {Polino}\ \emph {et~al.}(2020)\citenamefont {Polino},
  \citenamefont {Valeri}, \citenamefont {Spagnolo},\ and\ \citenamefont
  {Sciarrino}}]{Polino2020}%
  \BibitemOpen
  \bibfield  {author} {\bibinfo {author} {\bibfnamefont {E.}~\bibnamefont
  {Polino}}, \bibinfo {author} {\bibfnamefont {M.}~\bibnamefont {Valeri}},
  \bibinfo {author} {\bibfnamefont {N.}~\bibnamefont {Spagnolo}}, \ and\
  \bibinfo {author} {\bibfnamefont {F.}~\bibnamefont {Sciarrino}},\ }\href
  {\doibase 10.1116/5.0007577} {\bibfield  {journal} {\bibinfo  {journal} {AVS
  Quantum Science}\ }\textbf {\bibinfo {volume} {2}},\ \bibinfo {pages}
  {024703} (\bibinfo {year} {2020})}\BibitemShut {NoStop}%
\bibitem [{\citenamefont {Hassani}\ \emph {et~al.}(2017)\citenamefont
  {Hassani}, \citenamefont {Macchiavello},\ and\ \citenamefont
  {Maccone}}]{Hassani2017}%
  \BibitemOpen
  \bibfield  {author} {\bibinfo {author} {\bibfnamefont {M.}~\bibnamefont
  {Hassani}}, \bibinfo {author} {\bibfnamefont {C.}~\bibnamefont
  {Macchiavello}}, \ and\ \bibinfo {author} {\bibfnamefont {L.}~\bibnamefont
  {Maccone}},\ }\href {\doibase 10.1103/PhysRevLett.119.200502} {\bibfield
  {journal} {\bibinfo  {journal} {Phys. Rev. Lett.}\ }\textbf {\bibinfo
  {volume} {119}},\ \bibinfo {pages} {200502} (\bibinfo {year}
  {2017})}\BibitemShut {NoStop}%
\bibitem [{\citenamefont {Rezakhani}\ \emph {et~al.}(2019)\citenamefont
  {Rezakhani}, \citenamefont {Hassani},\ and\ \citenamefont
  {Alipour}}]{majid19}%
  \BibitemOpen
  \bibfield  {author} {\bibinfo {author} {\bibfnamefont {A.~T.}\ \bibnamefont
  {Rezakhani}}, \bibinfo {author} {\bibfnamefont {M.}~\bibnamefont {Hassani}},
  \ and\ \bibinfo {author} {\bibfnamefont {S.}~\bibnamefont {Alipour}},\ }\href
  {\doibase 10.1103/PhysRevA.100.032317} {\bibfield  {journal} {\bibinfo
  {journal} {Phys. Rev. A}\ }\textbf {\bibinfo {volume} {100}},\ \bibinfo
  {pages} {032317} (\bibinfo {year} {2019})}\BibitemShut {NoStop}%
\bibitem [{\citenamefont {Seveso}\ \emph {et~al.}(2019)\citenamefont {Seveso},
  \citenamefont {Albarelli}, \citenamefont {Genoni},\ and\ \citenamefont
  {Paris}}]{seveso19}%
  \BibitemOpen
  \bibfield  {author} {\bibinfo {author} {\bibfnamefont {L.}~\bibnamefont
  {Seveso}}, \bibinfo {author} {\bibfnamefont {F.}~\bibnamefont {Albarelli}},
  \bibinfo {author} {\bibfnamefont {M.~G.}\ \bibnamefont {Genoni}}, \ and\
  \bibinfo {author} {\bibfnamefont {M.~G.~A.}\ \bibnamefont {Paris}},\ }\href
  {\doibase 10.1088/1751-8121/ab599b} {\bibfield  {journal} {\bibinfo
  {journal} {Journal of Physics A: Mathematical and Theoretical}\ }\textbf
  {\bibinfo {volume} {53}},\ \bibinfo {pages} {02LT01} (\bibinfo {year}
  {2019})}\BibitemShut {NoStop}%
\bibitem [{\citenamefont {Feynman}(1986)}]{feynman1985quantum}%
  \BibitemOpen
  \bibfield  {author} {\bibinfo {author} {\bibfnamefont {R.~P.}\ \bibnamefont
  {Feynman}},\ }\href {\doibase 10.1007/BF01886518} {\bibfield  {journal}
  {\bibinfo  {journal} {Foundations of Physics}\ }\textbf {\bibinfo {volume}
  {16}},\ \bibinfo {pages} {507} (\bibinfo {year} {1986})}\BibitemShut
  {NoStop}%
\bibitem [{\citenamefont {Parthasarathy}(1988)}]{parthasarathy_1988}%
  \BibitemOpen
  \bibfield  {author} {\bibinfo {author} {\bibfnamefont {K.~R.}\ \bibnamefont
  {Parthasarathy}},\ }\href {\doibase 10.2307/3214153} {\bibfield  {journal}
  {\bibinfo  {journal} {Journal of Applied Probability}\ }\textbf {\bibinfo
  {volume} {25}},\ \bibinfo {pages} {151–166} (\bibinfo {year}
  {1988})}\BibitemShut {NoStop}%
\bibitem [{\citenamefont {Aharonov}\ \emph {et~al.}(1993)\citenamefont
  {Aharonov}, \citenamefont {Davidovich},\ and\ \citenamefont
  {Zagury}}]{AharonovPRA}%
  \BibitemOpen
  \bibfield  {author} {\bibinfo {author} {\bibfnamefont {Y.}~\bibnamefont
  {Aharonov}}, \bibinfo {author} {\bibfnamefont {L.}~\bibnamefont
  {Davidovich}}, \ and\ \bibinfo {author} {\bibfnamefont {N.}~\bibnamefont
  {Zagury}},\ }\href {\doibase 10.1103/PhysRevA.48.1687} {\bibfield  {journal}
  {\bibinfo  {journal} {Phys. Rev. A}\ }\textbf {\bibinfo {volume} {48}},\
  \bibinfo {pages} {1687} (\bibinfo {year} {1993})}\BibitemShut {NoStop}%
\bibitem [{\citenamefont {Kempe}(2003)}]{kempe2003}%
  \BibitemOpen
  \bibfield  {author} {\bibinfo {author} {\bibfnamefont {J.}~\bibnamefont
  {Kempe}},\ }\href {\doibase 10.1080/00107151031000110776} {\bibfield
  {journal} {\bibinfo  {journal} {Contemporary Physics}\ }\textbf {\bibinfo
  {volume} {44}},\ \bibinfo {pages} {307} (\bibinfo {year} {2003})}\BibitemShut
  {NoStop}%
\bibitem [{\citenamefont {Childs}\ and\ \citenamefont
  {Goldstone}(2004)}]{ChildsPRA}%
  \BibitemOpen
  \bibfield  {author} {\bibinfo {author} {\bibfnamefont {A.~M.}\ \bibnamefont
  {Childs}}\ and\ \bibinfo {author} {\bibfnamefont {J.}~\bibnamefont
  {Goldstone}},\ }\href {\doibase 10.1103/PhysRevA.70.022314} {\bibfield
  {journal} {\bibinfo  {journal} {Phys. Rev. A}\ }\textbf {\bibinfo {volume}
  {70}},\ \bibinfo {pages} {022314} (\bibinfo {year} {2004})}\BibitemShut
  {NoStop}%
\bibitem [{\citenamefont {Farhi}\ and\ \citenamefont
  {Gutmann}(1998)}]{Farhi98PRA}%
  \BibitemOpen
  \bibfield  {author} {\bibinfo {author} {\bibfnamefont {E.}~\bibnamefont
  {Farhi}}\ and\ \bibinfo {author} {\bibfnamefont {S.}~\bibnamefont
  {Gutmann}},\ }\href {\doibase 10.1103/PhysRevA.58.915} {\bibfield  {journal}
  {\bibinfo  {journal} {Phys. Rev. A}\ }\textbf {\bibinfo {volume} {58}},\
  \bibinfo {pages} {915} (\bibinfo {year} {1998})}\BibitemShut {NoStop}%
\bibitem [{\citenamefont {Ambainis}(2007)}]{ambainis2007}%
  \BibitemOpen
  \bibfield  {author} {\bibinfo {author} {\bibfnamefont {A.}~\bibnamefont
  {Ambainis}},\ }\href {https://doi.org/10.1137/S0097539705447311} {\bibfield
  {journal} {\bibinfo  {journal} {SIAM Journal on Computing}\ }\textbf
  {\bibinfo {volume} {37}},\ \bibinfo {pages} {210} (\bibinfo {year}
  {2007})}\BibitemShut {NoStop}%
\bibitem [{\citenamefont {Chandrashekar}\ \emph {et~al.}(2010)\citenamefont
  {Chandrashekar}, \citenamefont {Banerjee},\ and\ \citenamefont
  {Srikanth}}]{Chandrashekar10PRA}%
  \BibitemOpen
  \bibfield  {author} {\bibinfo {author} {\bibfnamefont {C.~M.}\ \bibnamefont
  {Chandrashekar}}, \bibinfo {author} {\bibfnamefont {S.}~\bibnamefont
  {Banerjee}}, \ and\ \bibinfo {author} {\bibfnamefont {R.}~\bibnamefont
  {Srikanth}},\ }\href {\doibase 10.1103/PhysRevA.81.062340} {\bibfield
  {journal} {\bibinfo  {journal} {Phys. Rev. A}\ }\textbf {\bibinfo {volume}
  {81}},\ \bibinfo {pages} {062340} (\bibinfo {year} {2010})}\BibitemShut
  {NoStop}%
\bibitem [{\citenamefont {Arrighi}\ \emph {et~al.}(2014)\citenamefont
  {Arrighi}, \citenamefont {Nesme},\ and\ \citenamefont
  {Forets}}]{Arrighi2014}%
  \BibitemOpen
  \bibfield  {author} {\bibinfo {author} {\bibfnamefont {P.}~\bibnamefont
  {Arrighi}}, \bibinfo {author} {\bibfnamefont {V.}~\bibnamefont {Nesme}}, \
  and\ \bibinfo {author} {\bibfnamefont {M.}~\bibnamefont {Forets}},\ }\href
  {\doibase 10.1088/1751-8113/47/46/465302} {\bibfield  {journal} {\bibinfo
  {journal} {Journal of Physics A: Mathematical and Theoretical}\ }\textbf
  {\bibinfo {volume} {47}},\ \bibinfo {pages} {465302} (\bibinfo {year}
  {2014})}\BibitemShut {NoStop}%
\bibitem [{\citenamefont {Mallick}\ and\ \citenamefont
  {Chandrashekar}(2016)}]{Chandrashekar16SR}%
  \BibitemOpen
  \bibfield  {author} {\bibinfo {author} {\bibfnamefont {A.}~\bibnamefont
  {Mallick}}\ and\ \bibinfo {author} {\bibfnamefont {C.~M.}\ \bibnamefont
  {Chandrashekar}},\ }\href {\doibase 10.1038/srep25779} {\bibfield  {journal}
  {\bibinfo  {journal} {Scientific Reports}\ }\textbf {\bibinfo {volume} {6}},\
  \bibinfo {pages} {25779} (\bibinfo {year} {2016})}\BibitemShut {NoStop}%
\bibitem [{\citenamefont {Arnault}\ and\ \citenamefont
  {Debbasch}(2016)}]{Arnault16PA}%
  \BibitemOpen
  \bibfield  {author} {\bibinfo {author} {\bibfnamefont {P.}~\bibnamefont
  {Arnault}}\ and\ \bibinfo {author} {\bibfnamefont {F.}~\bibnamefont
  {Debbasch}},\ }\href {\doibase https://doi.org/10.1016/j.physa.2015.08.011}
  {\bibfield  {journal} {\bibinfo  {journal} {Physica A: Statistical Mechanics
  and its Applications}\ }\textbf {\bibinfo {volume} {443}},\ \bibinfo {pages}
  {179} (\bibinfo {year} {2016})}\BibitemShut {NoStop}%
\bibitem [{\citenamefont {Mallick}\ \emph {et~al.}(2017)\citenamefont
  {Mallick}, \citenamefont {Mandal},\ and\ \citenamefont
  {Chandrashekar}}]{Mallick17}%
  \BibitemOpen
  \bibfield  {author} {\bibinfo {author} {\bibfnamefont {A.}~\bibnamefont
  {Mallick}}, \bibinfo {author} {\bibfnamefont {S.}~\bibnamefont {Mandal}}, \
  and\ \bibinfo {author} {\bibfnamefont {C.~M.}\ \bibnamefont
  {Chandrashekar}},\ }\href {\doibase 10.1140/epjc/s10052-017-4636-9}
  {\bibfield  {journal} {\bibinfo  {journal} {The European Physical Journal C}\
  }\textbf {\bibinfo {volume} {77}} (\bibinfo {year} {2017}),\
  10.1140/epjc/s10052-017-4636-9}\BibitemShut {NoStop}%
\bibitem [{\citenamefont {Arrighi}\ \emph {et~al.}(2018)\citenamefont
  {Arrighi}, \citenamefont {Di~Molfetta}, \citenamefont
  {M\'arquez-Mart\'{\i}n},\ and\ \citenamefont {P\'erez}}]{Arrighi18PRA}%
  \BibitemOpen
  \bibfield  {author} {\bibinfo {author} {\bibfnamefont {P.}~\bibnamefont
  {Arrighi}}, \bibinfo {author} {\bibfnamefont {G.}~\bibnamefont
  {Di~Molfetta}}, \bibinfo {author} {\bibfnamefont {I.}~\bibnamefont
  {M\'arquez-Mart\'{\i}n}}, \ and\ \bibinfo {author} {\bibfnamefont
  {A.}~\bibnamefont {P\'erez}},\ }\href {\doibase 10.1103/PhysRevA.97.062111}
  {\bibfield  {journal} {\bibinfo  {journal} {Phys. Rev. A}\ }\textbf {\bibinfo
  {volume} {97}},\ \bibinfo {pages} {062111} (\bibinfo {year}
  {2018})}\BibitemShut {NoStop}%
\bibitem [{\citenamefont {Arnault}\ \emph {et~al.}(2019)\citenamefont
  {Arnault}, \citenamefont {P\'erez}, \citenamefont {Arrighi},\ and\
  \citenamefont {Farrelly}}]{Arnault19PRA}%
  \BibitemOpen
  \bibfield  {author} {\bibinfo {author} {\bibfnamefont {P.}~\bibnamefont
  {Arnault}}, \bibinfo {author} {\bibfnamefont {A.}~\bibnamefont {P\'erez}},
  \bibinfo {author} {\bibfnamefont {P.}~\bibnamefont {Arrighi}}, \ and\
  \bibinfo {author} {\bibfnamefont {T.}~\bibnamefont {Farrelly}},\ }\href
  {\doibase 10.1103/PhysRevA.99.032110} {\bibfield  {journal} {\bibinfo
  {journal} {Phys. Rev. A}\ }\textbf {\bibinfo {volume} {99}},\ \bibinfo
  {pages} {032110} (\bibinfo {year} {2019})}\BibitemShut {NoStop}%
\bibitem [{\citenamefont {{De Vincenzo}}(2019)}]{DeVincenzo19RP}%
  \BibitemOpen
  \bibfield  {author} {\bibinfo {author} {\bibfnamefont {S.}~\bibnamefont {{De
  Vincenzo}}},\ }\href {\doibase https://doi.org/10.1016/j.rinp.2019.102598}
  {\bibfield  {journal} {\bibinfo  {journal} {Results in Physics}\ }\textbf
  {\bibinfo {volume} {15}},\ \bibinfo {pages} {102598} (\bibinfo {year}
  {2019})}\BibitemShut {NoStop}%
\bibitem [{\citenamefont {Singh}\ \emph {et~al.}(2019)\citenamefont {Singh},
  \citenamefont {Chandrashekar},\ and\ \citenamefont {Paris}}]{Singh}%
  \BibitemOpen
  \bibfield  {author} {\bibinfo {author} {\bibfnamefont {S.}~\bibnamefont
  {Singh}}, \bibinfo {author} {\bibfnamefont {C.~M.}\ \bibnamefont
  {Chandrashekar}}, \ and\ \bibinfo {author} {\bibfnamefont {M.~G.~A.}\
  \bibnamefont {Paris}},\ }\href {\doibase 10.1103/PhysRevA.99.052117}
  {\bibfield  {journal} {\bibinfo  {journal} {Phys. Rev. A}\ }\textbf {\bibinfo
  {volume} {99}},\ \bibinfo {pages} {052117} (\bibinfo {year}
  {2019})}\BibitemShut {NoStop}%
\bibitem [{\citenamefont {Zatelli}\ \emph {et~al.}(2020)\citenamefont
  {Zatelli}, \citenamefont {Benedetti},\ and\ \citenamefont
  {Paris}}]{Zatelli20}%
  \BibitemOpen
  \bibfield  {author} {\bibinfo {author} {\bibfnamefont {F.}~\bibnamefont
  {Zatelli}}, \bibinfo {author} {\bibfnamefont {C.}~\bibnamefont {Benedetti}},
  \ and\ \bibinfo {author} {\bibfnamefont {M.~G.~A.}\ \bibnamefont {Paris}},\
  }\href {\doibase 10.3390/e22111321} {\bibfield  {journal} {\bibinfo
  {journal} {Entropy}\ }\textbf {\bibinfo {volume} {22}} (\bibinfo {year}
  {2020}),\ 10.3390/e22111321}\BibitemShut {NoStop}%
\bibitem [{\citenamefont {Ragy}\ \emph {et~al.}(2016)\citenamefont {Ragy},
  \citenamefont {Jarzyna},\ and\ \citenamefont
  {Demkowicz-Dobrza\ifmmode~\acute{n}\else \'{n}\fi{}ski}}]{Ragy2016}%
  \BibitemOpen
  \bibfield  {author} {\bibinfo {author} {\bibfnamefont {S.}~\bibnamefont
  {Ragy}}, \bibinfo {author} {\bibfnamefont {M.}~\bibnamefont {Jarzyna}}, \
  and\ \bibinfo {author} {\bibfnamefont {R.}~\bibnamefont
  {Demkowicz-Dobrza\ifmmode~\acute{n}\else \'{n}\fi{}ski}},\ }\href {\doibase
  10.1103/PhysRevA.94.052108} {\bibfield  {journal} {\bibinfo  {journal} {Phys.
  Rev. A}\ }\textbf {\bibinfo {volume} {94}},\ \bibinfo {pages} {052108}
  (\bibinfo {year} {2016})}\BibitemShut {NoStop}%
\bibitem [{\citenamefont {Holevo}(1973)}]{Holevo1973}%
  \BibitemOpen
  \bibfield  {author} {\bibinfo {author} {\bibfnamefont {A.}~\bibnamefont
  {Holevo}},\ }\href {\doibase https://doi.org/10.1016/0047-259X(73)90028-6}
  {\bibfield  {journal} {\bibinfo  {journal} {Journal of Multivariate
  Analysis}\ }\textbf {\bibinfo {volume} {3}},\ \bibinfo {pages} {337 }
  (\bibinfo {year} {1973})}\BibitemShut {NoStop}%
\bibitem [{\citenamefont {Holevo}(1977)}]{Holevo1977}%
  \BibitemOpen
  \bibfield  {author} {\bibinfo {author} {\bibfnamefont {A.}~\bibnamefont
  {Holevo}},\ }\href {\doibase https://doi.org/10.1016/0034-4877(77)90009-X}
  {\bibfield  {journal} {\bibinfo  {journal} {Reports on Mathematical Physics}\
  }\textbf {\bibinfo {volume} {12}},\ \bibinfo {pages} {251 } (\bibinfo {year}
  {1977})}\BibitemShut {NoStop}%
\bibitem [{\citenamefont {Helstrom}(1976)}]{Helstrom1976}%
  \BibitemOpen
  \bibfield  {author} {\bibinfo {author} {\bibfnamefont {C.~W.}\ \bibnamefont
  {Helstrom}},\ }\href {\doibase https://doi.org/10.1007/BF01007479} {\emph
  {\bibinfo {title} {Quantum detection and estimation theory}}},\ Vol.~\bibinfo
  {volume} {3}\ (\bibinfo  {publisher} {Academic press New York},\ \bibinfo
  {year} {1976})\BibitemShut {NoStop}%
\bibitem [{\citenamefont {Holevo}(2011)}]{Holevo2011}%
  \BibitemOpen
  \bibfield  {author} {\bibinfo {author} {\bibfnamefont {A.~S.}\ \bibnamefont
  {Holevo}},\ }\href {https://www.springer.com/gp/book/9788876423758} {\emph
  {\bibinfo {title} {Probabilistic and statistical aspects of quantum
  theory}}},\ Vol.~\bibinfo {volume} {1}\ (\bibinfo  {publisher} {Springer
  Science \& Business Media},\ \bibinfo {year} {2011})\BibitemShut {NoStop}%
\bibitem [{\citenamefont {Helstrom}(1967)}]{Helstrom1967pla}%
  \BibitemOpen
  \bibfield  {author} {\bibinfo {author} {\bibfnamefont {C.}~\bibnamefont
  {Helstrom}},\ }\href {\doibase https://doi.org/10.1016/0375-9601(67)90366-0}
  {\bibfield  {journal} {\bibinfo  {journal} {Physics Letters A}\ }\textbf
  {\bibinfo {volume} {25}},\ \bibinfo {pages} {101 } (\bibinfo {year}
  {1967})}\BibitemShut {NoStop}%
\bibitem [{\citenamefont {Hayashi}\ and\ \citenamefont
  {Matsumoto}(2008)}]{Hayashi2008}%
  \BibitemOpen
  \bibfield  {author} {\bibinfo {author} {\bibfnamefont {M.}~\bibnamefont
  {Hayashi}}\ and\ \bibinfo {author} {\bibfnamefont {K.}~\bibnamefont
  {Matsumoto}},\ }\href {\doibase 10.1063/1.2988130} {\bibfield  {journal}
  {\bibinfo  {journal} {Journal of Mathematical Physics}\ }\textbf {\bibinfo
  {volume} {49}},\ \bibinfo {pages} {102101} (\bibinfo {year}
  {2008})}\BibitemShut {NoStop}%
\bibitem [{\citenamefont {Baumgratz}\ and\ \citenamefont
  {Datta}(2016{\natexlab{a}})}]{Baumgratz2016}%
  \BibitemOpen
  \bibfield  {author} {\bibinfo {author} {\bibfnamefont {T.}~\bibnamefont
  {Baumgratz}}\ and\ \bibinfo {author} {\bibfnamefont {A.}~\bibnamefont
  {Datta}},\ }\href {\doibase 10.1103/PhysRevLett.116.030801} {\bibfield
  {journal} {\bibinfo  {journal} {Phys. Rev. Lett.}\ }\textbf {\bibinfo
  {volume} {116}},\ \bibinfo {pages} {030801} (\bibinfo {year}
  {2016}{\natexlab{a}})}\BibitemShut {NoStop}%
\bibitem [{\citenamefont {Carollo}\ \emph {et~al.}(2019)\citenamefont
  {Carollo}, \citenamefont {Spagnolo}, \citenamefont {Dubkov},\ and\
  \citenamefont {Valenti}}]{Carollo2019}%
  \BibitemOpen
  \bibfield  {author} {\bibinfo {author} {\bibfnamefont {A.}~\bibnamefont
  {Carollo}}, \bibinfo {author} {\bibfnamefont {B.}~\bibnamefont {Spagnolo}},
  \bibinfo {author} {\bibfnamefont {A.~A.}\ \bibnamefont {Dubkov}}, \ and\
  \bibinfo {author} {\bibfnamefont {D.}~\bibnamefont {Valenti}},\ }\href
  {\doibase 10.1088/1742-5468/ab3ccb} {\bibfield  {journal} {\bibinfo
  {journal} {Journal of Statistical Mechanics: Theory and Experiment}\ }\textbf
  {\bibinfo {volume} {2019}},\ \bibinfo {pages} {094010} (\bibinfo {year}
  {2019})}\BibitemShut {NoStop}%
\bibitem [{\citenamefont {Albarelli}\ \emph {et~al.}(2019)\citenamefont
  {Albarelli}, \citenamefont {Friel},\ and\ \citenamefont
  {Datta}}]{Albarelli2019prl}%
  \BibitemOpen
  \bibfield  {author} {\bibinfo {author} {\bibfnamefont {F.}~\bibnamefont
  {Albarelli}}, \bibinfo {author} {\bibfnamefont {J.~F.}\ \bibnamefont
  {Friel}}, \ and\ \bibinfo {author} {\bibfnamefont {A.}~\bibnamefont
  {Datta}},\ }\href {\doibase 10.1103/PhysRevLett.123.200503} {\bibfield
  {journal} {\bibinfo  {journal} {Phys. Rev. Lett.}\ }\textbf {\bibinfo
  {volume} {123}},\ \bibinfo {pages} {200503} (\bibinfo {year}
  {2019})}\BibitemShut {NoStop}%
\bibitem [{\citenamefont {Albarelli}\ \emph {et~al.}(2020)\citenamefont
  {Albarelli}, \citenamefont {Barbieri}, \citenamefont {Genoni},\ and\
  \citenamefont {Gianani}}]{Albarelli2019}%
  \BibitemOpen
  \bibfield  {author} {\bibinfo {author} {\bibfnamefont {F.}~\bibnamefont
  {Albarelli}}, \bibinfo {author} {\bibfnamefont {M.}~\bibnamefont {Barbieri}},
  \bibinfo {author} {\bibfnamefont {M.}~\bibnamefont {Genoni}}, \ and\ \bibinfo
  {author} {\bibfnamefont {I.}~\bibnamefont {Gianani}},\ }\href {\doibase
  https://doi.org/10.1016/j.physleta.2020.126311} {\bibfield  {journal}
  {\bibinfo  {journal} {Physics Letters A}\ }\textbf {\bibinfo {volume}
  {384}},\ \bibinfo {pages} {126311} (\bibinfo {year} {2020})}\BibitemShut
  {NoStop}%
\bibitem [{\citenamefont {Demkowicz-Dobrza{\'{n}}ski}\ \emph
  {et~al.}(2020)\citenamefont {Demkowicz-Dobrza{\'{n}}ski}, \citenamefont
  {G{\'{o}}recki},\ and\ \citenamefont {Gu{\c{t}}{\u{a}}}}]{Demko2020}%
  \BibitemOpen
  \bibfield  {author} {\bibinfo {author} {\bibfnamefont {R.}~\bibnamefont
  {Demkowicz-Dobrza{\'{n}}ski}}, \bibinfo {author} {\bibfnamefont
  {W.}~\bibnamefont {G{\'{o}}recki}}, \ and\ \bibinfo {author} {\bibfnamefont
  {M.}~\bibnamefont {Gu{\c{t}}{\u{a}}}},\ }\href {\doibase
  10.1088/1751-8121/ab8ef3} {\bibfield  {journal} {\bibinfo  {journal} {Journal
  of Physics A: Mathematical and Theoretical}\ }\textbf {\bibinfo {volume}
  {53}},\ \bibinfo {pages} {363001} (\bibinfo {year} {2020})}\BibitemShut
  {NoStop}%
\bibitem [{\citenamefont {Razavian}\ \emph {et~al.}(2020)\citenamefont
  {Razavian}, \citenamefont {Paris},\ and\ \citenamefont
  {Genoni}}]{Razavian2020}%
  \BibitemOpen
  \bibfield  {author} {\bibinfo {author} {\bibfnamefont {S.}~\bibnamefont
  {Razavian}}, \bibinfo {author} {\bibfnamefont {M.~G.}\ \bibnamefont {Paris}},
  \ and\ \bibinfo {author} {\bibfnamefont {M.~G.}\ \bibnamefont {Genoni}},\
  }\href {\doibase 10.3390/e22111197} {\bibfield  {journal} {\bibinfo
  {journal} {Entropy}\ }\textbf {\bibinfo {volume} {22}},\ \bibinfo {pages}
  {1197} (\bibinfo {year} {2020})}\BibitemShut {NoStop}%
\bibitem [{\citenamefont {Fiderer}\ \emph {et~al.}(2021)\citenamefont
  {Fiderer}, \citenamefont {Tufarelli}, \citenamefont {Piano},\ and\
  \citenamefont {Adesso}}]{FidererPRX}%
  \BibitemOpen
  \bibfield  {author} {\bibinfo {author} {\bibfnamefont {L.~J.}\ \bibnamefont
  {Fiderer}}, \bibinfo {author} {\bibfnamefont {T.}~\bibnamefont {Tufarelli}},
  \bibinfo {author} {\bibfnamefont {S.}~\bibnamefont {Piano}}, \ and\ \bibinfo
  {author} {\bibfnamefont {G.}~\bibnamefont {Adesso}},\ }\href {\doibase
  10.1103/PRXQuantum.2.020308} {\bibfield  {journal} {\bibinfo  {journal} {PRX
  Quantum}\ }\textbf {\bibinfo {volume} {2}},\ \bibinfo {pages} {020308}
  (\bibinfo {year} {2021})}\BibitemShut {NoStop}%
\bibitem [{\citenamefont {Nayak}\ and\ \citenamefont
  {Vishwanath}(2000)}]{nayak2000}%
  \BibitemOpen
  \bibfield  {author} {\bibinfo {author} {\bibfnamefont {A.}~\bibnamefont
  {Nayak}}\ and\ \bibinfo {author} {\bibfnamefont {A.}~\bibnamefont
  {Vishwanath}},\ }\href {https://arxiv.org/abs/quant-ph/0010117} {\bibfield
  {journal} {\bibinfo  {journal} {arXiv preprint quant-ph/0010117}\ } (\bibinfo
  {year} {2000})}\BibitemShut {NoStop}%
\bibitem [{\citenamefont {Brun}\ \emph {et~al.}(2003)\citenamefont {Brun},
  \citenamefont {Carteret},\ and\ \citenamefont {Ambainis}}]{Brun2003}%
  \BibitemOpen
  \bibfield  {author} {\bibinfo {author} {\bibfnamefont {T.~A.}\ \bibnamefont
  {Brun}}, \bibinfo {author} {\bibfnamefont {H.~A.}\ \bibnamefont {Carteret}},
  \ and\ \bibinfo {author} {\bibfnamefont {A.}~\bibnamefont {Ambainis}},\
  }\href {\doibase 10.1103/PhysRevA.67.032304} {\bibfield  {journal} {\bibinfo
  {journal} {Phys. Rev. A}\ }\textbf {\bibinfo {volume} {67}},\ \bibinfo
  {pages} {032304} (\bibinfo {year} {2003})}\BibitemShut {NoStop}%
\bibitem [{\citenamefont {Annabestani}\ \emph {et~al.}(2010)\citenamefont
  {Annabestani}, \citenamefont {Akhtarshenas},\ and\ \citenamefont
  {Abolhassani}}]{Mostafa2010}%
  \BibitemOpen
  \bibfield  {author} {\bibinfo {author} {\bibfnamefont {M.}~\bibnamefont
  {Annabestani}}, \bibinfo {author} {\bibfnamefont {S.~J.}\ \bibnamefont
  {Akhtarshenas}}, \ and\ \bibinfo {author} {\bibfnamefont {M.~R.}\
  \bibnamefont {Abolhassani}},\ }\href {\doibase 10.1103/PhysRevA.81.032321}
  {\bibfield  {journal} {\bibinfo  {journal} {Phys. Rev. A}\ }\textbf {\bibinfo
  {volume} {81}},\ \bibinfo {pages} {032321} (\bibinfo {year}
  {2010})}\BibitemShut {NoStop}%
\bibitem [{\citenamefont {Baumgratz}\ and\ \citenamefont
  {Datta}(2016{\natexlab{b}})}]{Baum2016}%
  \BibitemOpen
  \bibfield  {author} {\bibinfo {author} {\bibfnamefont {T.}~\bibnamefont
  {Baumgratz}}\ and\ \bibinfo {author} {\bibfnamefont {A.}~\bibnamefont
  {Datta}},\ }\href {\doibase 10.1103/PhysRevLett.116.030801} {\bibfield
  {journal} {\bibinfo  {journal} {Phys. Rev. Lett.}\ }\textbf {\bibinfo
  {volume} {116}},\ \bibinfo {pages} {030801} (\bibinfo {year}
  {2016}{\natexlab{b}})}\BibitemShut {NoStop}%
\bibitem [{\citenamefont {Apellaniz}\ \emph {et~al.}(2018)\citenamefont
  {Apellaniz}, \citenamefont {Urizar-Lanz}, \citenamefont {Zimbor\'as},
  \citenamefont {Hyllus},\ and\ \citenamefont {T\'oth}}]{Apellaniz2018}%
  \BibitemOpen
  \bibfield  {author} {\bibinfo {author} {\bibfnamefont {I.}~\bibnamefont
  {Apellaniz}}, \bibinfo {author} {\bibfnamefont {I.~n.}\ \bibnamefont
  {Urizar-Lanz}}, \bibinfo {author} {\bibfnamefont {Z.}~\bibnamefont
  {Zimbor\'as}}, \bibinfo {author} {\bibfnamefont {P.}~\bibnamefont {Hyllus}},
  \ and\ \bibinfo {author} {\bibfnamefont {G.}~\bibnamefont {T\'oth}},\ }\href
  {\doibase 10.1103/PhysRevA.97.053603} {\bibfield  {journal} {\bibinfo
  {journal} {Phys. Rev. A}\ }\textbf {\bibinfo {volume} {97}},\ \bibinfo
  {pages} {053603} (\bibinfo {year} {2018})}\BibitemShut {NoStop}%
\end{thebibliography}%
\end{document}